\documentclass[12pt]{article}
\usepackage{epsfig}
\usepackage{a4wide}

\usepackage{latexsym}
\usepackage{amssymb}
\usepackage{amsmath}
\usepackage{authblk}

\usepackage{graphics}
\usepackage{authblk}
\usepackage{graphicx}
\usepackage{float}

\usepackage{natbib}
\setcitestyle{authoryear,open={(},close={)}}

\title
{Effect of small floating disks on the propagation of gravity waves}

\author{F. De Santi$^{1,2}$ \& P. Olla$^{1,2}$}
\affil{\small{
$^1$ISAC-CNR, Sez. Cagliari, I--09042 Monserrato, Italy 
\\
$^2$INFN, Sez. Cagliari, I--09042 Monserrato, Italy 
}
}

\numberwithin{equation}{section}
\begin{document}

\def\beq{\begin{equation}}
\def\eeq{\end{equation}}
\def\d{{\rm d}}
\def\im{{\rm i}}
\def\e{{\bf e}}
\def\ex{{\rm e}}
\def\Rfr{R_{{\scriptscriptstyle\rm Fr}}}
\def\LambdaFu{\Lambda_{{\scriptscriptstyle\rm Fu}}}
\def\Rfl{R_{{\scriptscriptstyle\rm Fl}}}
\def\sd{\strut\displaystyle}
\def\diss{\bar\epsilon_{{\scriptscriptstyle T}}}
\def\bOmega{{\boldsymbol{\Omega}}}
\def\bsigma{{\boldsymbol{\sigma}}}
\def\homega{{\hat\omega}}
\def\r{{\bf r}}
\def\hatrho{{\hat\varrho}}
\def\bp{{\bar p}}
\def\p{{\bf p}}
\def\g{{\bf g}}
\def\A{{\bf A}}
\def\C{{\bf C}}
\def\f{{\bf f}}
\def\u{{\bf u}}
\def\w{{\bf w}}
\def\U{{\bf U}}
\def\V{{\bf V}}
\def\PhiU{\Phi^{\scriptscriptstyle U}}
\def\AU{A^{\scriptscriptstyle U}}
\def\cm{{\rm cm}}
\def\l{{\bf l}}
\def\sec{{\rm s}}
\def\Ckol{C_{Kol}}
\def\flux{\bar\epsilon}
\def\latent{{\cal L}}
\def\kdeep{k_{\scriptscriptstyle \infty}}
\def\qHh{q_{\scriptscriptstyle H-h}}
\def\qH{q_{\scriptscriptstyle H}}
\def\kH{k_{\scriptscriptstyle H}}
\def\hkH{\hat k_{\scriptscriptstyle H}}
\def\smali{{\scriptscriptstyle i}}
\def\smalfi{{\scriptscriptstyle \frac{5}{3} }}
\def\smalL{{\scriptscriptstyle{\rm L}}}
\def\smalP{{\scriptscriptstyle {\rm P}}}
\def\smalT{{\scriptscriptstyle {\rm T}}}
\def\smalE{{\scriptscriptstyle{\rm E}}}
\def\smal1n{{\scriptscriptstyle (1.1,n)}}
\def\smaln{{\scriptscriptstyle (n)}}
\def\smalA{{\scriptscriptstyle {\rm A}}}
\def\smalze{{\scriptscriptstyle (0)}}
\def\smalun{{\scriptscriptstyle (1)}}
\def\smalzeze{{\scriptscriptstyle (0,0)}}
\def\smalzeun{{\scriptscriptstyle (0,1)}}
\def\smalzedu{{\scriptscriptstyle (0,2)}}
\def\smalzen{{\scriptscriptstyle (0,n)}}
\def\unmezz{{\scriptscriptstyle 1/2}}
\def\smaldu{{\scriptscriptstyle (2)}}
\def\smaltr{{\scriptscriptstyle (3)}}
\def\smaln{{\scriptscriptstyle (n)}}
\def\smalel{{\scriptscriptstyle l}}
\def\gammaP{\gamma^\smalP}
\def\shell{{\tt S}}
\def\ball{{\tt B}}
\def\nav{\bar N}
\def\micron{\mu{\rm m}}
\font\brm=cmr10 at 24truept
\font\bfm=cmbx10 at 15truept

\maketitle

\begin{abstract}
A dispersion relation for gravity waves in water covered by disk-like impurities embedded in a
viscous matrix is derived. The macroscopic equations are obtained by ensemble-averaging 
the fluid equations at the disk scale in the asymptotic limit of long waves and low disk surface 
fraction. Various regimes are identified
depending on the disk radii and the thickness and viscosity of the top layer.
Semi-quantitative analysis in the close-packing regime suggests
dramatic modification of the dynamics, with orders of magnitude
increase in wave damping and wave dispersion.
A simplified model working in this regime is proposed.
Possible applications to 
wave propagation in ice-covered ocean is discussed and comparison with field
data is provided.
\vfill\eject

\end{abstract}

\section{Introduction}\label{Introduction}
Materials floating on the surface of the ocean affect the propagation of gravity waves:
oil slicks have long been known to produce wave attenuation.
Larger objects, such as floating debris, buoys,
ice floes, have a more complicated effect, but attenuation is
usually dominant. 

Analogous effects are commonly observed in polar regions when sea ice is present.
Remote sensing of wave propagation modifications 
has been used as a proxy for the thickness of newly formed ice
\citep{wadhams02,wadhams04,squire08,doble15}.
The matter is of great interest in climate modelling, 
as sea ice contributes in important ways to moderate the global climate.
It is conceivable that similar approaches 
find application in oil spill detection \citep{brekke05}.

Over the years, several models of wave propagation in ice covered waters
have been proposed \citep{squire95,squire07}. Such models depend crucially
on the properties of the ice, which in turn depend on its age.
The initial phase of ice formation is characterized by so called grease ice, that is a
a thick suspension of ice crystals (frazil ice).
The crystals coalesce to form cake-shaped
objects (pancake ice), which initially have 
diameter 30--100 cm and thickness 10--30 cm, and later
evolve into floes several meters wide. 

In the case of grease ice, one of the first theories of wave propagation was derived 
by \cite{weber87}, where the suspension was treated as a very viscous medium in
creeping flow conditions. \cite{keller98} extended the theory
to generic values of the viscosity. Various generalizations have been proposed
to include the effect of an eddy viscosity in the otherwise inviscid bottom region
\citep{decarolis02}, the possibility of a viscoelastic
component in the ice \citep{wang10} and spatial inhomogeneities \citep{wang11}.
For additional references and a comparison of different 
viscoelastic models (see \cite{mosig15}).

In the case of large floes, the floe-wave and floe-floe interaction could be modelled
as a wave scattering process \citep{foldy45,bennetts09}, with the flexural
dynamics of the individual floes expected to play a dominant role
\citep{meylan02,kohout08,wadhams73}. In the opposite limit of pancakes, which are
much smaller than a wavelength, scattering and elastic properties are not expected to be
important. Rather, viscous forces from the grease ice and collisions should dominate.
A macroscopic model, in which the ice layer is treated as a continuum with 
assigned rheological properties, seems therefore natural. 

Macroscopic models, unfortunately, depend on rheological parameters such
as effective viscosities and elastic moduli, which 
must be supplied 
either from experiments or by fitting field data. Moreover, there is no guarantee that 
such parameterizations properly account for the physics of the
problem. 

A first attempt to derive rheological properties 
from the microscopic dynamics was presented, in the case of grease ice, by \cite{decarolis05}.
One wonders whether a similar approach
could be used with pancake ice, by treating the pancakes as microscopic
on the scale of the waves.
As in the case of grease ice, the main
difficulty lies in the fact that one is dealing with a concentrated suspension,
as pancakes typically form a closely packed assembly at the ocean surface.

A possible strategy in analysis of the problem is to consider first the dilute limit, and to
use the information gathered in this way to get some insight of the behavior of the
system in a concentrated condition. This is precisely the approach that will be
followed in the present paper. 

We consider first the simpler problem of the dynamics of
a monodisperse two-dimensional suspension of non-interacting thin disks in the
field of a gravity wave. To mimick real pancake ice, we assume the
disks to be embedded in a viscous matrix (the grease ice layer) lying on top
of an inviscid fluid column.
The stress modifications at the water surface are determined as
an average effect from the flow perturbation by the individual pancakes.
At a macroscopic scale, this takes the form of modified boundary conditions 
on the wave field at the water surface. Such boundary conditions
may be interpreted equivalently
as a spatially uniform three-layer model, with an infinitely
thin top layer accounting for the effect of the pancakes. 

We shall use this information to derive a semiquantitative model of the disk
dynamics in the close-packing regime.
The reduced relative mobility of the disks with respect to
the dilute case is expected to cause a sharp increase of the friction forces
on the grease ice matrix. We shall provide order of magnitude estimates
of such forces and incorporate them in the wave dispersion relation.
The resulting modification in wave propagation will be compared with
the prediction by the dilute theory.


This paper is organized as follows.
In Sec. \ref{Flow}, the flow perturbation by the wave field around an isolated disk
is considered.  In Sec. \ref{Stress}, a coarse graining operation 
is carried out to evaluate the average stress generated locally in the wave field.
In Sec. \ref{Dispersion}, the resulting modification to the dispersion relation
is determined.  In Sec. \ref{Relevance}, some qualitative considerations
on the close-packing limit is presented.  In Sec. \ref{Discussion} the results are discussed
and compared with other models.
Section \ref{Conclusion} is devoted to conclusions.
Calculation details are confined to the Appendices.

\section{Flow perturbation by a single disk}
\label{Flow}
Consider a random distribution of disks of radius $R$
and thickness $\delta\ll R$, floating on top of an infinitely deep column of
fluid of viscosity $\nu$ and density $\varrho$. We postpone analysis of the case in which 
only the top part of the column is viscous to Sec. \ref{Two-layer}.
We assume that a small amplitude 
gravity wave of frequency $\omega$ is propagating in 
the fluid. We want to determine the response of the disks to the
wave field in the dilute limit, in which no interaction among the
disks is present. 

The problem is characterized by two relevant space scales. 
One is induced by
the wavenumber in the case of an infinitely deep 
inviscid fluid (without the disks), $\kdeep=\omega^2/g$, with
$g\simeq 9.8\ {\rm m^2/s}$ 
the gravitational acceleration. The other is the thickness of the viscous boundary layer 
at the water surface, 
\beq
\lambda_\alpha=(\nu/\omega)^{1/2}, 
\label{lambda_alpha}
\eeq
that is the momentum diffusion length in a wave period \citep{longuet53}. 
For waves of unperturbed wavelength $\lambda\approx 100$ m, we 
would have  $\omega\approx 0.78$ rad/s. A typical estimate for the grease ice viscosity is
$\nu\approx 0.01\ {\rm m^2/s}$ \citep{newyear99,wadhams04}. This would
produce a boundary layer of thickness $\lambda_\alpha\approx 0.1$ m. Shorter waves
would produce even thinner boundary layers. Smallness of this parameter is an illustration
that 
creeping flow assumptions, characteristic of
standard suspension theory do not apply at the disk scale.  

We assume the ordering
\beq
\delta,\lambda_\alpha\ll R\ll \kdeep^{-1}
\label{geometric constraint}
\eeq
and introduce expansion parameters
\beq
\epsilon_k=\kdeep R\quad{\rm and}\quad
\epsilon_\alpha=\lambda_\alpha/R.
\label{expansion parameters}
\eeq
Presence of an isolated disk will affect the wave in substantially two ways:
\begin{itemize}
\item
Possible relative motion of the disk with respect to the fluid.
\item
Fluid stress at the disk surface due to the rigid structure of the body.
\end{itemize}
The first is basically an inertia effect, which is going
to be negligible for very thin disks.  To evaluate
the second effect, we must calculate the flow perturbation generated by interaction of the
disk with the wave field.

Let us put our reference frame with origin at the disk center, with the $z$-axis pointing 
upward and the $x$-axis in the direction of propagation of the wave. 
For small amplitude
waves, the velocity field at the water surface can be approximated with that at
the unperturbed water surface $z=0$. For small $kx$, we
can Taylor expand the velocity field in the absence of the disk,
\beq
\U(\r,t)=\U(0,t)+x\partial_x\U(0,t)+\frac{1}{2}x^2\partial_x^2\U(0,t)+\ldots
\label{expansion}
\eeq
The disk will experience a tangential stress
proportional to $\partial_xU_x$ associated with extension and compression in the $x$ direction,
and a normal stress proportional to $\partial_x^2U_z$ associated with bending. 
If $\delta/R\ll 1$, the disk will have very low inertia so that its relative motion with 
respect to the fluid will be negligible. For small $\epsilon_k$, the disk will thus
translate with velocity $U_{disk,x}(t)\simeq U_x(0,t)$ and rotate with angular frequency
$\bOmega_{disk}(t)\simeq-\partial_xU_z(0,t)\,\e_y$. 
The difference $U_{disk,z}(t)-U_z(0,t)$ must be determined explicitly by 
equating to zero the total normal force on the disk. Analysis carried out in Sect. 
\ref{Potential} will demonstrate that $U_{disk,z}(t)- U_z(0,t)$ is small.

We consider rigid disks.
No-slip and impermeability condition must be imposed. For small $\delta/R$, 
the boundary conditions need to be enforced only at the disk bottom,
$z=0$, $\rho=\sqrt{x^2+y^2}<R$. The velocity perturbation at the disk surface will be
\beq
\u(\r,t)=\U_{disk}(t)+\Omega_{disk}\ x\,\e_z-\U(\r,t).
\eeq
Exploiting Eq. (\ref{expansion}), we obtain
\beq
\u(\r,t)=2a (x/R)\e_x+[b+2c (x/R)^2]\e_z,\quad z=0,\ \rho<R,
\label{perturb}
\eeq
where $a=-(R/2)\partial_xU_x(0,t)$, $c=-(R/4)\partial_x^2U_z(0,t)$, and  $b$ 
gives the relative vertical motion of the disk with respect to the fluid.

It is convenient to shift to cylindrical coordinates
and express the velocity as a sum over angular harmonics. 
We write for the generic quantity
${\cal Q}$,
\beq
{\cal Q}(\r,t)=\sum_{m=-\infty}^{+\infty}{\cal Q}_m(\rho,z;t)\ex^{\im m\phi}.
\label{harmonics}
\eeq
The boundary condition for $\rho<R$, Eq. (\ref{perturb}), will read in cylindrical coordinates
\begin{eqnarray}
\u_0(\rho,0)=a (\rho/R)\e_\rho+\Big[b+c(\rho/R)^2\Big]\e_z;
\nonumber
\\
\u_{\pm 2}(\rho,0)=(1/2)\Big[a(\rho/R)\e_\rho \pm\im a(\rho/R)\e_\phi+c(\rho/R)^2\e_z\Big].
\label{abc}
\end{eqnarray}
For $\rho>R$, we have to impose zero stress at the free water surface,
\begin{eqnarray}
\tau_{z\phi}=\mu(\partial_zu_\rho+\partial_\rho u_z)=0,
\qquad
\tau_{z\rho}=\mu(\partial_zu_\phi+\frac{1}{\rho}\partial_\phi u_z)=0,
\nonumber
\\
\tau_{zz}=
2\mu\partial_zu_z-P=0,\quad z=0,\ \rho>R,
\label{tangential stress}
\end{eqnarray}
where $P$ is the pressure perturbation and $\mu=\varrho\nu$ is the dynamic
viscosity of the fluid.

The velocity perturbation $\u$ obeys, for small-amplitude waves, 
the time-dependent Stokes equation
\beq
\partial_t\u+\varrho^{-1}\nabla (P+V)=\nu\nabla^2\u,\qquad \nabla\cdot\u=0,
\label{time-dep-Stokes}
\eeq
where $V=-\varrho gz$ is the gravitational potential.
We can express $\u$ in terms of scalar and vector potentials
\beq
\u=-\nabla\Phi+\nabla\times\A.
\label{potentials}
\eeq
In angular components:
\begin{eqnarray}
\u_m&=&\Big(-\partial_\rho\Phi_m+\frac{\im m}{\rho}A_{m,z}
-\partial_zA_{m,\phi}\Big)\e_\rho
\nonumber
\\
&+&\Big(-\frac{\im m}{\rho}\Phi_m+\partial_zA_{m,\rho}-\partial_\rho A_{m,z}\Big)\e_\phi
\nonumber
\\
&+&\Big(-\partial_z\Phi_m+\partial_\rho A_{m,\phi}+\frac{1}{\rho }(A_{m,\phi}
-\im mA_{m,\rho })\Big)\e_z.
\label{velocity}
\end{eqnarray}
Note that for $m=0$ we can take $\A_0=A_0\e_\phi$ 
(the flow component for $m=0$ is in essence two-dimensional).

The scalar and vector potentials
$\Phi$ and $\A$ can be taken to obey, from Eq. (\ref{time-dep-Stokes}),
\beq
\varrho\partial_t\Phi=P+V,\qquad
\nabla^2\Phi=0
\label{potential pressure}
\eeq
and
\beq
\partial_t\A=\nu\nabla^2\A,\qquad\nabla\cdot\A=0
\label{potential-equations}
\eeq
(see Appendix \ref{Potential representation}).
The first of Eq. (\ref{potential pressure}) 
can be used to rewrite the condition of
zero normal stress at $\rho >R$, Eq. (\ref{tangential stress}), in the form
\beq
2\nu\partial_zu_z+g\eta_z-\partial_t\Phi=0,
\label{normal stress}
\eeq
where $\eta_z$ is the vertical displacement of the water surface induced
by $\u$. We have the kinematic relation
\beq
\dot\eta_z(\rho ,\phi;t)=u_z(\rho ,\phi,0;t).
\label{eta}
\eeq 
The system of equations formed by the second of Eq. (\ref{potential pressure}) and
(\ref{potential-equations}), with the definition
Eq. (\ref{potentials}) and the boundary conditions Eqs. (\ref{tangential stress})
and (\ref{normal stress}), describes the dynamics of the flow
perturbation induced by the disk. 
For $\nu\to 0$, the velocity field $\u(\r,t)$ 
describes the flow that would be produced (in the absence of the wave) 
by a radius $R$ membrane whose surface
oscillates vertically with the law  $u_z(\r,t)=b+2c(x/R)^2$. 
For $\kdeep R\to 0$, the
effect would be that of a point force quadrupole. Inclusion of viscosity 
induces local dissipation, which we shall evaluate perturbatively in the limit of small
$\epsilon_\alpha$ and $\epsilon_k$.

\subsection{Boundary layer structure}
\label{Boundary}
The small $\epsilon_\alpha$ limit is associated with a viscous boundary layer 
asymptotically thin on the scale of the disk. This suggests a multiscale approach to 
calculate the vector potential 
\beq
\A(\r)=\A_+(\r)\ex^{\alpha z},
\label{Ahat}
\eeq
with
\beq
\alpha=(-\im\omega/\nu)^{1/2}
\eeq
identifying the fast scale
and $\A_+$ slowly dependent on $z$.
We set up the perturbation expansion 
\beq
\Phi=\sum_{n=0}^{+\infty}\Phi^\smaln\epsilon_\alpha^n,\qquad
\A=\sum_{n=1}^{+\infty}
\A^\smaln\epsilon_\alpha^n,
\eeq
and use
Eq. (\ref{velocity}) to
write the boundary conditions 
(\ref{tangential stress}) and (\ref{normal stress}) in terms of potentials.
For $\epsilon_\alpha\ll 1$, we expect distinct behaviours of $\A$ for $\rho <R$ and $\rho >R$,
separated by a transition region of thickness $\lambda_\alpha\ll R$.
Keeping only leading order terms, we have in the inner region $\rho <R$,
from Eq. (\ref{velocity}): 
\begin{eqnarray}
&&\partial_z\Phi^\smalze_m=-u^\smalze_{m,z},
\qquad
\bar\alpha A^\smalun_{m,\phi}=-u_{m,\rho }-\partial_\rho \Phi^\smalze_m,
\nonumber
\\
&&\bar\alpha A^\smalun_{m,\rho }=u_{m,\phi}+\frac{\im m}{\rho}\Phi_m^\smalze,
\qquad\bar\alpha=\alpha/\epsilon_\alpha
\label{internal}
\end{eqnarray}
(note that $u_{m,z}$ depends on $b$ and must be determined within perturbation theory).
The zero divergence condition for $\A$ becomes, for $r<R$:
\beq
A^\smalun_{m,z}=0;\qquad
\frac{\im m}{\rho }A^\smalun_{m,\phi}+\partial_\rho A^\smalun_{m,\rho }
+\frac{1}{\rho }A^\smalun_{m,\rho }+\bar\alpha A^\smaldu_{m,z}=0.
\label{divergenceless internal}
\eeq
In the outer region $r>R$, we get, from
Eqs. (\ref{tangential stress}) and (\ref{normal stress}),
$A^\smalun_{m,\rho }=A^\smalun_{m,\phi}=0$, which gives to leading order:
\begin{eqnarray}
-\partial_\rho \partial_z\Phi^\smalze_m+\frac{\im m\bar\alpha}{\rho }A_{m,z}^\smalun
-\bar\alpha^2A_{m,\phi}^\smaldu=0,
\nonumber
\\
-\frac{2\im m}{\rho }\partial_z\Phi_m^\smalze-\bar\alpha\partial_\rho A_{m,z}^\smalun
+\bar\alpha^2A_{m,\rho }^\smaldu=0,
\nonumber
\\
-g\eta_{z,m}^\smalze+\partial_t\Phi_m^\smalze=0
\label{external}
\end{eqnarray}
(it is easy to see that $\nu\partial_zu_z/\partial_t\Phi=O(\epsilon_\alpha^2)$, 
while $V_m$ and $\partial_t\Phi$ are of the same order in $\epsilon_\alpha$).
The zero divergence condition for $\A$ becomes, for $\rho >R$:
\beq
A_{m,z}^\smalun=0;
\qquad
\frac{\im m}{\rho }A^\smaldu_{m,\phi}+\partial_\rho A^\smaldu_{m,\rho }
+\frac{1}{\rho }A^\smaldu_{m,\rho }+\bar\alpha A^\smaltr_{m,z}=0.
\label{divergenceless external}
\eeq

Putting together Eqs. 
(\ref{internal}) to (\ref{divergenceless external}), 
we get to lowest order in $\epsilon_\alpha$
the boundary conditions at $z=0$:
\begin{eqnarray}
\partial_z\Phi^\smalze_m=-u^\smalze_{m,z},
\qquad
\bar\alpha A^\smalun_{m,\phi}=-u_{m,\rho }-\partial_\rho \Phi^\smalze_m,
\nonumber
\\
\bar\alpha A^\smalun_{m,\rho }=u_{m,\phi}+\frac{\im m}{\rho }\Phi^\smalze_m,\quad 
A^\smalun_{m,z}=0,
\qquad \rho<R,
\label{boundary conditions in}
\end{eqnarray}
and
\beq
\A_m^\smalun=0, 
\qquad -g\eta^\smalze_{z,m}+\partial_t\Phi^\smalze_m=0,
\qquad
\rho >R,
\label{boundary conditions out}
\eeq
The divergenceless condition Eq. (\ref{divergenceless internal})
ceases to be necessary (it would provide us with the second order term 
$A^\smaldu_{m,z}$ that we do not need at the order considered).
Similarly, the zero tangential stress conditions at $\rho >R$ is also automatically satisfied
at the order considered.

\subsection{Potential component}
\label{Potential}
The potential component of the flow is fully accounted for by the part of the velocity
field due to the scalar potential $\Phi$. This obeys the Laplace equation $\nabla^2\Phi=0$,
with the boundary conditions established by the first of Eq. (\ref{boundary conditions in})
and the second of Eq. (\ref{boundary conditions out}).
The boundary condition $g\eta^\smalze_{z,m}+\partial_t\Phi^\smalze_m=0$ 
in the outer region $\rho >R$, can 
be rewritten in terms of potentials using Eqs. (\ref{normal stress}) and (\ref{eta}). From 
Eqs. (\ref{velocity}) and the first of Eq.  (\ref{boundary conditions out}), we have
in the external region $\rho >R$, to lowest order in $\epsilon_\alpha$,
$u_z=-\partial_z\Phi^\smalze$. 
Putting together with the first of Eq. (\ref{boundary conditions in}),
we get the boundary conditions for the scalar potential:
\begin{eqnarray}
\partial_z\Phi^\smalze=-u^\smalze_z,\qquad\qquad \rho <R,
\nonumber
\\
\partial_z\Phi^\smalze-\frac{\omega^2}{g}\Phi^\smalze=0,\qquad \rho >R,
\label{boundary conditions Phi}
\end{eqnarray}
i.e. mixed Neumann and Robin boundary conditions. These must be compounded with the
condition of zero vertical force on the disk,
required to fix the parameter $b$ in Eq. (\ref{perturb}). This is
\beq
\int_{\rho <R}\d S\ \Big[u^\smalze_z(\r)-\frac{\omega^2}{g}\Phi^\smalze(\r)\Big]=0,
\label{zero force}
\eeq
where the integral is carried out on the disk surface at $z=0$.

We solve the boundary value problem defined by Eqs. (\ref{boundary conditions Phi}) and
(\ref{zero force})  perturbatively in $\epsilon_k$ and to lowest order in 
$\epsilon_\alpha$. We write
\beq
\Phi^\smalze=\sum_{n=0}^{+\infty}\Phi^\smalzen\epsilon_k^n,
\eeq
and similarly for $b$ and $u_z$.
It is easy to see that small $\epsilon_k$ corresponds to a condition of slow dynamics
for the potential component of the flow. This means again that inertia is negligible
at the disk scale, which converts the Robin  boundary condition at 
$\rho >R$ in Eq. (\ref{boundary conditions Phi}),
to lowest order in $\epsilon_k$, to a Neumann boundary condition 
$\partial_z\Phi^\smalzeze=0$. 

We recall the expression for the Neumann Green function for the Laplace equation
(see e.g. 
\cite{jackson}):
\beq
G^N(\r,\hat\r)=\frac{1}{|\r-\hat\r|}+\frac{1}{|\r-\hat\r'|},
\qquad
\hat\r'=(\hat x,\hat y,-\hat z),
\label{Neumann}
\eeq
which allows us to write
\beq
\Phi^\smalzeze(\r)=
-\frac{1}{2\pi}\int_{\rho_0<R}\d S_0\ \frac{u^\smalzeze_z(\rho_0)}{|\r-\r_0|}.
\label{Phi^smalze}
\eeq
In similar way, the total vertical force in Eq. (\ref{zero force}) will receive
contribution, to lowest order, only from the vertical velocity. In other words,
the condition of zero vertical force on the disk coincides with that of zero
average vertical component of the velocity perturbation.
This gives in Eq. (\ref{perturb})
\beq
b^\smalzeze=-c.
\label{b^smalze}
\eeq
The next orders in the expansion are obtained in iterative fashion from the expression for
$\Phi^\smalzeun$
\begin{eqnarray}
\partial_z\Phi^\smalzeun=-b^\smalzeun,\qquad\ \  \rho <R,
\nonumber
\\
\partial_z\Phi^\smalzeun=-\frac{1}{R}\Phi^\smalzeze,\qquad \rho >R,
\label{boundary conditions Phi^smalun}
\end{eqnarray}
where $b^\smalzeun$ is obtained from the next order in the condition of zero average normal force,
Eq. (\ref{zero force}):
\beq
b^\smalzeun=\frac{1}{\pi R^2}\int_{\rho <R}\d S\ \Phi^\smalzeze(\r).
\label{b^smalun}
\eeq
The coefficient $b^\smalzeun$ gives the first contribution to the relative vertical motion of 
the disk with respect to the fluid. 
From now on we shall neglect subscripts on $\Phi$ and $\A$, and indicate
$\Phi\simeq\Phi^\smalzeze$, $\A\simeq\epsilon_k\A^\smalun=(A_\rho,A_\phi,0)$.


\section{The stress perturbation}
\label{Stress}
We want to determine the stress generated on the water surface by the disks. In the
dilute limit, this is the sum of the stresses generated by the disks individually,
neglecting their mutual interaction.
By construction, the only place where the surface stress is non-zero is under a disk. From Eq.
(\ref{velocity}), the stress under a disk will be, 
working to lowest order in $\epsilon_k$ and $\epsilon_\alpha$:
\begin{eqnarray}
\tau_{m,z\rho }&=&\mu\Big(\partial_zu_{m,\rho }+\partial_\rho u_{m,z}\Big)\simeq
-\mu\alpha^2A_{m,\phi},
\label{papera01}
\\
\tau_{m,z\phi}&=&\mu\Big(\partial_zu_{m,\phi}+\frac{1}{\rho }\partial_\phi u_{m,z}\Big)\simeq
\mu\alpha^2A_{m,\rho },
\label{papera02}
\\
\tau_{m,zz}&=&2\mu\partial_zu_{m,z}+V_m-\varrho\partial_t\Phi_m
\simeq
\frac{\im \varrho g}{\omega}u_{m,z}.
\label{papera03}
\end{eqnarray}
We note that 
the vector potential in the tangential components can be expressed
by means of Eq.  (\ref{boundary conditions in}), as a function of the velocity
$\u$ and of derivatives of the scalar potential $\Phi$. 
Thus, the only field whose spatial structure we actually 
need to know is the scalar potential $\Phi$.

At macroscopic scale, the cumulative effects of the disks is evaluated by means
of a local spatial average, which is carried out by summing over all the possible positions
of a disk, relative to a hypothetical fixed sensor.

If the disks are distributed randomly, uniformly on the water surface,
the only stress components surviving the average will be, by symmetry, the ones along
$xz$ and $zz$. We find from Eqs. (\ref{papera01}) and (\ref{papera02}),
\beq
\langle\tau_{zx}\rangle=-\frac{f\alpha^2\mu}{\pi R^2}
\int_0^{2\pi}\d\phi\int_0^R \rho\,\d \rho\ 
\Big[A_\phi\cos\phi+A_\rho\sin\phi\Big],
\label{papera1}
\eeq
while, from Eq. (\ref{papera03}),
\beq
\langle\tau_{zz}\rangle=\frac{\im f\varrho g}{\pi R^2\omega}
\int_0^{2\pi}\d\phi\int_0^R \rho\,\d \rho\ u_z;
\label{papera2}
\eeq
$f$ is the surface fraction of the disks, which
represents the probability that a disk actually lies over the sensor.

It is clear that the integrals in Eqs. (\ref{papera1}) and (\ref{papera2}) can be carried out 
equivalently in the disk reference frame, by summing over the sensor positions. 
This allows us to use the expressions for  the integrands in the previous sections.
Care must be taken, however, of the fact that
the expansion in Eq. (\ref{perturb}) is now carried out with respect to different
positions in the wave field.

\begin{figure}
\begin{center}
\includegraphics[width=5.5cm]{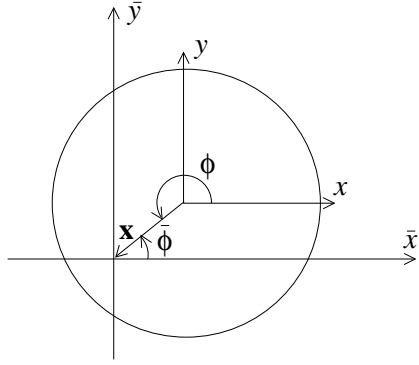}
\caption{Laboratory and floating disk reference frames.}
\label{panfig1}
\end{center}
\end{figure}

Let us indicate with $\bar\r=(\bar x,\bar y,0)=(\bar\rho,\bar\phi,0)$ the position of the disk
in the laboratory reference frame, and place the sensor at $\bar\r=0$
(see Fig. \ref{panfig1}).
In the disk reference frame, the sensor will be at $\r=-\bar\r$,
where $\r=(\rho,\phi,0)=(\bar r,\bar\phi-\pi,0)$, and the velocity $\U(\r,t)$ at
the sensor position will be related to the corresponding expression in the
laboratory frame, $\bar\U(0,t)$, by
\beq
\bar\U(0,t)=\U(\r,t)=\bar\U(\bar\r,t)+x\partial_{\bar x}\bar\U(\bar\r,t)+
\frac{1}{2}x^2\partial^2_{\bar x}\bar\U(\bar\r,t)+\ldots
\label{Taylor}
\eeq
This gives us the dependence of the coefficients $a$ and $c$ in Eq. 
(\ref{perturb}), on the sensor position $\r=-\bar\r$:
\begin{eqnarray}
a(\r,t)&=&a(0,t)+\frac{Rx}{2}\partial_{\bar x}^2\bar U_x(\bar\r,t)|_{\bar\r=0}+\ldots,
\nonumber
\\
c(\r,t)&=&c(0,t)+\frac{R^2x}{4}\partial_{\bar x}^3\bar U_z(\bar\r,t)|_{\bar\r=0}
-\frac{R^2x^2}{8}\partial_{\bar x}^4\bar U_z(\bar\r,t)|_{\bar\r=0}+\ldots,
\label{papera3}
\end{eqnarray}
where $a(0,t)$ and $c(0,t)$ are the values of $a$ and $c$ when the disk center
is at the sensor position. It is important to note that neglecting the corrections
in Eq. (\ref{papera3}) would give zero in Eqs. (\ref{papera1}) and
(\ref{papera2}), as the lowest order contribution to the average stress is just 
the total force on the disk---which is zero, divided by the disk area.

We are now in the position to calculate the average stress. Let us start with the
tangential stress. Working to lowest order in 
$\epsilon_k$ and $\epsilon_\alpha$, we have,
from Eqs. (\ref{papera1}), (\ref{boundary conditions in}) and (\ref{perturb}):
\begin{eqnarray}
\langle\tau_{xz}\rangle&=&
\frac{f\mu\alpha}{\pi R^2}\int_0^{2\pi}\d\phi\int_0^R\rho\,\d\rho\ 
\Big[\Big(u_\rho+\partial_\rho\Phi\Big)\cos\phi
-\Big(u_\phi+\frac{1}{\rho}\partial_\phi\Phi\Big)\sin\phi\Big].
\nonumber
\\
&=&\frac{f\mu\alpha}{\pi R^2}\int_0^{2\pi}\d\phi\int_0^R\rho\,\d\rho\
\Big[\Big(a\frac{\rho}{R}+\partial_\rho\Phi_0\Big)\cos\phi
\nonumber
\\
&+&\Big(a\frac{\rho}{R}+ 2\partial_\rho\Phi_2\Big)
\cos 2\phi\cos\phi+\Big(a\frac{\rho}{R}+\frac{4}{\rho}\Phi_2\Big)\sin 2\phi\sin\phi\Big].
\label{papera4}
\end{eqnarray}
We stress that the coefficient $a$ (and $c$ through $\Phi_{0,2}$) depend on $\r$ 
through Eq. (\ref{papera3}).

Calculations, detailed in Appendix \ref{Green}, allow to write the potential harmonics
$\Phi_0$ and $\Phi_2$
in terms of the corresponding harmonics of the Green function $G^N$  appearing
in Eq. (\ref{Neumann}). Substituting the expansion in Eq. (\ref{papera3}) into 
Eq. (\ref{papera4}), gives, after additional algebra:
\beq
\langle\tau_{xz}\rangle=
\frac{11f\mu R^2\alpha}{64}\frac{\partial^2\bar U_x}{\partial\bar x^2}
-\frac{Bf\mu\alpha R^3}{2}\frac{\partial^3\bar U_z}{\partial\bar x^3},
\label{pi_xz}
\eeq
where $B\simeq 0.16$ (see Eq. (\ref{B})). 
From comparison of  Eqs. 
(\ref{boundary conditions in}), 
(\ref{papera3}) and (\ref{papera4}), it is
clear that the first term to RHS of Eq. (\ref{pi_xz}) accounts for the $u_{r,\phi}$
contributions to $\langle\tau_{xz}\rangle$ while the second accounts for the one
by $\Phi_{0,2}$.

Passing to analysis of the normal stress, 
substituting Eqs. (\ref{velocity}) and 
(\ref{papera3}) into Eq. (\ref{papera2}) will give
\beq
\langle\tau_{zz}\rangle
=\frac{\im f\varrho gR^4}{64\omega}\frac{\partial^4\bar U_z}{\partial\bar x^4}.
\label{pi_zz}
\eeq
Despite the higher derivatives with respect to $\bar x$ in $\tau_{zz}$, the two stress
component are of the same order,
\beq
\frac{\langle\tau_{zz}\rangle}{\langle\tau_{xz}\rangle}\sim
\frac{\epsilon_k}{\epsilon_\alpha}.
\label{xi}
\eeq
On the contrary, the second term to RHS of Eq. (\ref{pi_xz}) is smaller than the first
by a factor $\epsilon_k$ and should be disregarded to the order considered. In the
end, to lowest order
in $\epsilon_k$ and $\epsilon_\alpha$,
neither component of the stress depends on the spatial structure of $\Phi$.

\subsection{The case of a finite thickness viscous layer}
\label{Two-layer}
We can extend the analysis to the case in which 
only a top layer of thickness $h$ of the water column is viscous, and
the whole basin (including the viscous layer on top) has finite depth $H$. 
We assume
\beq
H\gg R,\qquad
\kdeep h\ll 1,
\eeq
and take the difference $\varrho_w-\varrho$ between the densities in the inviscid and viscous
regions to be small and positive.

Let us consider the modification to the flow perturbation by a single disk.
Viscous stresses are generated only in the top part of the column at $-h<z<0$, while
the flow remains potential in the bottom part $-H<z<-h$. 
The vector potential, which is now confined to the top viscous layer, in order
to insure continuity of tangential stress at $z=-h$, will thus acquire an
additional component
\beq
\A(\r)=\hat\A^+(\r)\ex^{\alpha z}+\hat\A^-(\r)\ex^{-\alpha z}.
\label{papera6}
\eeq
The spatial structure of $\Phi$ will similarly be modified by the the solid boundary 
at $z=-H$ and by the discontinuities in $\tau_{zz}$ and $(\nabla\times\A)_z$ at $z=-h$. 

We want to understand how all this affects the boundary conditions at $z=0$.

Consider first the normal stress.
Inspection of Eq. (\ref{papera03}) tells us
that, to lowest order in $\epsilon_{\alpha,k}$, 
the normal stress is determined solely by the velocity condition on $u_z$,
and is insensitive to the spatial structure of $\Phi$ 
(the only place in which  the $\epsilon_k\ll 1$ assumption plays a role
is the Taylor expansion in Eqs. (\ref{expansion}) and (\ref{perturb})). The
normal stress at the surface thus remains unaffected by presence of a rigid bottom at $z=-H$
and of a viscous-inviscid transition at $z=-h$.

Let us shift our attention to the tangential stress.
We can decompose
$\hat\A^-=-\hat\A^+\ex^{-2\alpha h}+\tilde\A$, where $-\hat\A^+\ex^{-2\alpha h}$
cancels the tangential stress contribution from $\hat\A^+$, and $\tilde\A$
cancels the one from  $\Phi$. To evaluate $\tilde\A$, we
must determine the two contributions to stress from $\Phi$ and $\A$, that we indicate with
$\tau_{\Phi,A}$.
From Eq. (\ref{velocity}) we obtain 
$\tau_\Phi|_{z=0}\approx \mu\Phi|_{z=0}/R^2$ and $\tau_A|_{z=0}\approx \mu\alpha^2 A|_{z=0}$,
which gives $\tau_\Phi|_{z=0}\approx \epsilon_\alpha^2 \tau_A|_{z=0}$.
We have at most $\tau_\Phi|_{z=-h}\approx \epsilon_\alpha^2 \tau_A|_{z=0}$, so that
$\tilde \A\lesssim \epsilon_\alpha^2 A|_{z=0}\ex^{-\alpha h}\approx 
\epsilon_\alpha^2\hat A^+\ex^{-\alpha h}$. 

When $h\gg\lambda_\alpha$, both $\hat A^-$ and
the corresponding modification to the boundary condition at $z=0$ are 
exponentially small. When $h\approx\lambda_\alpha$, 
the contribution from $\tilde A$ to $\hat A^-$ is 
$O(\epsilon_\alpha^2)$ and can
be disregarded. We can thus write in general
\beq
\hat\A^-=-\hat\A^+\ex^{-2\alpha h}.
\eeq
From here we obtain for 
the tangential stress at $z=0$, exploiting Eq. (\ref{boundary conditions in}):
\beq
\tau_{z\rho}= \mu\alpha u_{m,\rho}\tanh(\hat\alpha\psi)
\quad{\rm and}\quad
\tau_{z\phi}= \mu\alpha u_{m,\phi}\tanh(\hat\alpha\psi),
\label{modification2}
\eeq
where we have introduced dimensionless quantities
\beq
\psi=\frac{h}{\lambda_\alpha}=\frac{h\kdeep^{1/4}g^{1/4}}{\nu^{1/2}}
\qquad{\rm and}\qquad
\hat\alpha=\alpha\lambda_\alpha=\sqrt{-\im}.
\label{psi}
\eeq 

We arrive at the general expression for the average stress at the surface, working
to lowest order in $\epsilon_k$ and $\epsilon_\alpha$:
\beq
\langle\tau_{xz}\rangle=\zeta\alpha\partial_{\bar x}^2\bar U_x,
\qquad
\langle\tau_{zz}\rangle=\frac{\im \sigma}{\omega}\partial_{\bar x}^4\bar U_z,
\label{modification3}
\eeq
where, from Eqs. (\ref{pi_xz}) and (\ref{pi_zz}),
\beq
\zeta=\frac{11f\mu R^2}{64}\tanh(\hat\alpha\psi)\quad{\rm and}\quad
\sigma=\frac{fg\varrho R^4}{64}.
\label{zeta sigma}
\eeq
We see that the disk layer acts as a membrane with bending rigidity
$\sigma$ and extensional viscosity $\alpha\zeta$. We note the dependence of $\sigma$ on 
an exogenous variable such as $g$, and the complex nature and frequency dependence of 
$\zeta$, which cannot be easily interpreted in terms of a viscoelastic dynamics
such as the one described by \cite{wang10}.

\section{Dispersion relation}
\label{Dispersion}
The procedure to derive a dispersion relation for gravity waves in the presence of a viscous
layer at the surface is analogous to the one described in 
\citep{keller98,decarolis02,wang11}.
We have to enforce four boundary conditions: continuity of tangential and normal 
stress at the water surface, $z=0$; zero tangential stress at the bottom of the
viscous layer, $z=-h$;  continuity of normal stress again at $z=-h$.
Addition of the disks generates non-zero surface stresses, as accounted for by
Eqs. (\ref{modification3}) and (\ref{zeta sigma}). Imposing continuity between 
the fluid and the surface stresses
gives us:
\begin{eqnarray}
&&\mu(\partial_x U_z+\partial_z U_x)=\zeta\alpha\partial_x^2 U_x,
\label{tg}
\\
&&2\mu\partial_z U_z-P=\frac{\im \sigma}{\omega}\partial_x^4 U_z
\label{nrm}
\end{eqnarray}
(we omit from now on overbars on vectors in the laboratory frame).
We write the velocity field of the wave in terms of
potentials: $ U_x=-\partial_x\PhiU-\partial_z \AU$,
$U_z=-\partial_z\PhiU+\partial_x \AU$. 
In the top viscous layer $-h<z<0$, we have
from Eq. (\ref{time-dep-Stokes}):
\begin{eqnarray}
\PhiU=\PhiU_+\ex^{kz+\im(kx-\omega t)}+\PhiU_-\ex^{-kz+\im(kx-\omega t)},
\nonumber
\\
 \AU= \AU_+\ex^{\alpha_k z+\im(kx-\omega t)}+ \AU_-\ex^{-\alpha_k z+\im(kx-\omega t)},
\end{eqnarray}
where
$\alpha_k=(-\im\omega/\nu+k^2)^{1/2}$. In the inviscid region
$-H<z<-h$, only the scalar potential survives:
\beq
\PhiU=\PhiU_w\cosh[k(z+H)]\ex^{\im(kx-\omega t)},
\eeq
where we have enforced the zero vertical velocity condition at the bottom of the column,
$z=-H$.

It is convenient to introduce dimensionless quantities
\begin{eqnarray}
&&\hat k=\frac{k}{\kdeep};
\quad
\hat\nu=\frac{\kdeep^{3/2}\nu}{g^{1/2}};
\quad
\hat\alpha_k=\sqrt{-\im+\hat\nu\hat k^2};\quad
\quad 
\hat h=\kdeep  h;
\quad \hat H=\kdeep H;
\nonumber
\\
&& 
\hatrho=\frac{\varrho}{\varrho_w};\quad
\xi=\frac{\epsilon_k}{\epsilon_\alpha}=\frac{\kdeep^{5/4}g^{1/4}R^2}{\nu^{1/2}};\quad
\hat\zeta=\frac{11\xi f}{64}\tanh(\hat\alpha\psi);\quad
\hat\sigma=\frac{\xi^2f}{64}.
\label{dimensionless}
\end{eqnarray}
Dependence on the two small parameters $\epsilon_k$ and $\epsilon_\alpha$
has been replaced by one on $\hat\nu\equiv (\epsilon_k\epsilon_\alpha)^2$
and $\hat h$. 
For wavelength $\approx 100$ m, effective viscosity $\nu\approx 0.01\ {\rm m^2/s}$ and 
thickness of the viscous layer $h\approx 0.5$ m, we would have $\psi\approx 4.4$, corresponding
to $\hat\nu\approx 5\cdot 10^{-5}$ and $\hat h\approx 0.03$. 
Note that we can write $\hat h=\hat\nu^{1/2}\psi$, and
since $\psi$ is $O(1)$ in most situations of interest, we end up with a
single small parameter $\hat\nu=(\kdeep\lambda_\alpha)^2$, which is independent
of $R$. 
The relevant parameter accounting
for the disk radius is now $\xi$, that in the case of pancake ice tends to 
be rather small (with the same
wave parameters as before, taking $R\approx 0.5$ m would give $\xi\approx 0.14$), but
could become larger than one for lower viscosity and shorter waves.

In terms of potentials, the continuity condition for the tangential stress at the surface,
Eq. (\ref{tg}), becomes
\begin{eqnarray}
&&\hat\nu\hat k^2[(2+\hat\zeta\hat\alpha\hat k)\PhiU_+
+(-2+\hat\zeta\hat\alpha\hat k)\PhiU_-]
-(1+\hat\nu^{1/2}\hat\zeta\hat k^2+2\im\hat\nu\hat k^2) \AU_+
\nonumber
\\
&&-(1-\hat\nu^{1/2}\hat\zeta\hat k^2+2\im\hat\nu\hat k^2) \AU_-=0.
\label{eq1}
\end{eqnarray}
In similar way, the continuity condition on the surface normal stress, Eq.  (\ref{nrm}), 
becomes, using Eqs. (\ref{potential pressure}), (\ref{normal stress}) and (\ref{eta}) to express
pressure in terms of potentials:
\begin{eqnarray}
&&[\hat k-1-\hat\nu\hat k^2(2\im+\hat\sigma\hat k^3)]\PhiU_+
+[-\hat k-1-\hat\nu\hat k^2(2\im-\hat\sigma\hat k^3)]\PhiU_-
\nonumber
\\
&&-\im\hat k(1-2\im\hat\nu^{1/2}\hat\alpha_k-\hat\nu\hat\sigma\hat k^4) \AU_+
-\im\hat k(1+2\im\hat\nu^{1/2}\hat\alpha_k-\hat\nu\hat\sigma\hat k^4) \AU_-=0.
\label{eq2}
\end{eqnarray}
We see that
for $\xi$ fixed, sending $\hat\nu$ to zero corresponds to sending 
to zero also the contribution from the disks
(the $\hat\nu\to 0$ limit at fixed $\xi$ coincides with 
an $R\to 0$ limit at fixed $\kdeep$ and $\nu$).

Calculations analogous to those leading to Eqs. (\ref{eq1}) and (\ref{eq2})
allow us to write continuity conditions at the
interface for the tangential stress:
\begin{eqnarray}
&&2\hat\nu\hat k^2(\PhiU_+\ex^{-\hat k\hat h}-
\PhiU_-\ex^{\hat k\hat h})
-(1+2\im\hat\nu\hat k^2)( \AU_+\ex^{-\hat\alpha_k\psi}
+ \AU_-\ex^{\hat\alpha_k\psi})=0,
\label{eq3}
\end{eqnarray}
and for the normal stress
(see Appendix \ref{Boundary conditions}):
\begin{eqnarray}
&&\{\im[\hatrho-\qHh+(1-\hatrho)\hat k]-2\hatrho\hat\nu\hat k^2\}\PhiU_+\ex^{-\hat k\hat h}
+\{\im[\hatrho+\qHh-(1-\hatrho)\hat k]-2\hatrho\hat\nu\hat k^2\}\PhiU_-\ex^{\hat k\hat h}
\nonumber
\\
&&+[(1-\hatrho)\hat k-\qHh+2\im\hatrho\hat\nu^{1/2}\hat\alpha_k\hat k] \AU_+
\ex^{-\hat\alpha_k\psi}
\nonumber
\\
&&+[(1-\hatrho)\hat k-\qHh-2\im\hatrho\hat\nu^{1/2}\hat\alpha_k\hat k] \AU_-
\ex^{\hat\alpha_k\psi}
=0,
\label{eq4}
\end{eqnarray}
where 
\beq
\qH(\hat k)=\frac{1}{\tanh(\hat k\hat H)}.
\label{qH}
\eeq

We have a system of four equations (\ref{eq1}), (\ref{eq2}), (\ref{eq3}) and (\ref{eq4}), in
the four variables $\PhiU_\pm$ and $ \AU_\pm$, which, 
for $\hat\zeta=\hat\sigma=0$, reduce to Eqs. (15-18) in \cite{keller98}.
From here, a dispersion relation can 
be extracted equating to zero the secular determinant.
We proceed perturbatively in $\hat\nu^{1/2}$ or equivalently, for $\psi$ not large and fixed,
in powers of $\hat h$.
We write
\beq
\hat k=\sum_{n=0}^{+\infty}\hat k^\smaln\hat h^n
\eeq
and likewise expand the secular determinant
\beq
S(\hat k,\hat h)=\Big[S
+\hat h(\hat k^\smalun\partial_{\hat k}+\partial_{\hat h})S+
\ldots\Big]_{\hat k=\hat k^\smalze,\hat h=0}.
\label{secular}
\eeq
The dispersion relation $S(\hat k,\hat h)=0$ is 
solved equating to zero order by order the coefficients in the expansion in
Eq. (\ref{secular}).
The operation is sped-up with the help of a symbolic manipulation program. 


Let us focus for the moment on the case of an infinitely deep basin,
$H\to\infty$, for which $\hat k^\smalze=1$. 
Stopping the perturbative expansion at $O(\hat\nu^{3/2})$, we obtain 
\begin{eqnarray}
\hat k\simeq 1+\hat\nu\hatrho[\im\hat\alpha\hat\zeta+\hat\sigma]+
\hat\nu^{3/2}\psi\Big\{
8\hatrho\Big[\im+\frac{\hat\alpha}{\psi}\frac{\cosh\hat\alpha\psi-1}{\sinh\hat\alpha\psi}\Big]
\nonumber
\\
+2\hatrho(1-\hatrho)\hat\sigma+\Big[2\hat\alpha\hatrho(1-\hatrho)-\frac{4\hatrho}{\psi}
\frac{\cosh\hat\alpha\psi-1}{\sinh\hat\alpha\psi}\Big]\im\hat\zeta
-\frac{\hatrho\hat\alpha\cosh\hat\alpha\psi}{\psi\sinh\hat\alpha\psi}\hat\zeta^2\Big\},
\label{reldisp3}
\end{eqnarray}
which has a number of relevant limit regimes.

\subsection{Limit regimes}
If the surface fraction of the disks
is not too small and the viscous layer is not too thin, 
\beq
\hat k\simeq 1+\hatrho\hat\nu[\im\hat\alpha\hat\zeta+\hat\sigma],
\qquad f,\psi,\xi\quad {\rm finite.}
\label{basic}
\eeq
Writing $\im\hat\alpha=2^{-1/2}(1+\im)$, we see that
disks produce
a frequency-dependent response consisting of both wave damping and decreased
wave propagation speed.
The viscous layer contributes only a correction at $O(\hat\nu^{3/2})$. 
The information on the
layer depth is buried in the dependence on $\psi$ of the tangential stress $\hat\zeta$.

The limit of a very thin viscous layer, $h\ll\lambda_\alpha$, which corresponds to putting
$\psi\ll 1$ in Eq. (\ref{reldisp3}), gives the result, from the first of Eq. (\ref{zeta sigma}):
\beq
\hat k\simeq 1+\hatrho\hat\nu\hat\sigma,\qquad \psi\quad{\rm small}.
\label{shallow}
\eeq
Also in this case, the leading contribution comes from 
the disks.\footnote{No dissipation to this order, as
the flow perturbation from the disks, in the absence of a viscous layer, becomes 
purely potential.}

The viscous layer will play a role
in the absence of disks, i.e. for $f\to 0$, 
or when the disks are small, i.e. for $\xi\to 0$.
We get in this case
\beq
\hat k\simeq 1+8\hatrho\hat\nu^{3/2}
\Big[\im\psi+\hat\alpha\frac{\cosh\hat\alpha\psi-1}{\sinh\hat\alpha\psi}\Big],
\quad f\ {\rm or}\ \xi\ \ {\rm small},
\label{vis}
\eeq
which can be brought back to
the small $\hat h$ deep-water limit of the dispersion relation, Eq. (45)
in \cite{wang10}; see also \citep{keller98}.

Finally, the limit of an infinitely deep viscous layer 
could be obtained converting the perturbation 
expansion in powers of $\hat\nu$ at fixed $\psi$, to one in powers of $\hat\nu$ at fixed
$\hat h$. The result, stopping at $O(\hat\nu)$, is
\beq
\hat k\simeq 1+\hat\nu
\frac{[4(1-\ex^{-2\hat h})+\hat\alpha\hat\zeta]\im+\hat\sigma}
{1+(1/\hatrho-1)\ex^{-2\hat h}},\qquad \hat h\ {\rm finite},
\label{order-two}
\eeq
which, for large $\hat h$, becomes
\beq
\hat k=1+\hat\nu[4\im+\im\hat\alpha\hat\zeta+\hat\sigma],\qquad\hat h\to\infty,
\label{reldisp}
\eeq
and we recognize, in the disk-free case $f=0$, the dispersion relation for waves in a
viscous fluid derived by \cite{lamb}. 
The transition 
from a shallow to a deep layer regime occurs in two stages. For $\psi\ll 1$, the disks 
see the viscous layer as shallow, corresponding to the dispersion relation in Eq. (\ref{shallow}). 
They will see the layer as deep for $\psi\gg 1$, which would correspond to setting
$\tanh(\hat\alpha\psi)=1$ in the first of Eq. (\ref{zeta sigma}) and then in Eq. (\ref{basic}).
Only for $\hat h\gg 1$, will Eq. (\ref{order-two}) converge to the infinite depth 
solution Eq. (\ref{reldisp}), and will the waves see the viscous layer infinitely deep. 


\section{Close-packing effects}
\label{Relevance}
\begin{figure}
\begin{center}
\includegraphics[width=5.5cm]{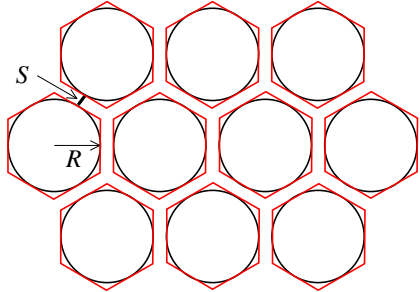}
\caption{
Sketch of close-packing arrangement. The maximum area fraction $f_{max}$
is obtained for
$S=0$, and coincides with the ratio of the circle to hexagon area:
$f_{max}=\pi/(2\sqrt{3})\simeq 0.91$. The deviation $f_{max}-f$ is the surface
fraction of the interstices among the hexagons,
which, for $S\ll R$, is $\propto S/R$.}
\label{panfig2}
\end{center}
\end{figure}

Moving away from the dilute limit requires taking into consideration the mutual interaction
of the disks. A microscopic theory generalizing Eqs. (\ref{modification3}) and (\ref{zeta sigma})
is going to be  difficult. Some progress can be made if we assume that 
the disks interaction is produced by contact forces. To fix the ideas let us imagine
that the disks are arranged in a regular lattice, as illustrated in 
Fig. \ref{panfig2}.
The wave velocity field at the surface, $U_x(x,t)$, is horizontally compressible.
The disk separation $S$ will thus oscillate in space and time, and, if $S$ is initially small, 
collisions will occur.\footnote{Since disk inertia is small, collisions can be treated as
anelastic. At the same time, the kinetic energy dissipated in collisions is neglected
compared to the viscous dissipation at the disk bottom.}
Such collisions take place in the compression regions where
$\partial_xU_x(x,t)<0$. If rafting is neglected, the disks will remain locked in their
position relative to neighbours until $\partial_xU_x(x,t)$ becomes positive again. 
In such compression regions the wave will not see the disks as individual entities, rather
as horizontally rigid agglomerates (``islands'') 
whose extension $\Delta$ scales with the wavelength $\lambda$.

We can try to be more quantitative on this.

The wave field at $z=0$,
\beq
U_x(x,t)=\frac{\mathcal{A}g}{\omega}\cos(kx-\omega t+\varphi),
\eeq
where $2\mathcal{A}g/\omega^2$ is the crest to trough wave height, determines the motion of the
disks.
The relative motion of a pair of points separated by $X\ll\lambda$ in the
$x$ direction obeys
$\dot X\simeq X\partial_xU_x(x,t)$, which can be integrated to give
\beq
X(t)\simeq X(0)[1+\mathcal{A}\cos(kx-\omega t+\varphi)].
\label{eta_x}
\eeq
Consider two disks aligned along $x$, whose centers are separated initially
by $X(0)\approx 2R$.
The maximum relative displacement in a wave period will be $\approx R\mathcal{A}$ and
the minimum rim to rim separation between
neighboring disks, compatible with horizontal free motion in the wave field,
will thus be $S\approx R\mathcal{A}$ (see Fig.  \ref{panfig2}).

Collisions among disks will take place in the regions of the wave in which
\beq
\mathcal{A}\cos(kx-\omega t+\varphi)>S/R
\label{condition}
\eeq
(see Eq. (\ref{eta_x})). These regions are centered at the coordinates
of instantaneous maximum compression $x_l=k^{-1}[\omega t-\varphi+(2l+1)\pi]$.
Since in Eq. (\ref{condition})  $S/R\approx f_{max}-f$, the extension of these regions
is independent of $R$ and scales with $\lambda$ as claimed,
\beq
\Delta=\frac{4}{\pi}\sqrt{\frac{\gamma}{11}}\lambda,
\qquad
\gamma=\gamma\Big(\frac{f_{max}-f}{\cal{A}}\Big),
\label{gamma}
\eeq
where the proportionality coefficient is chosen
to simplify the dispersion relation to be derived below.
In the limit $f\to f_{max}$, we can imagine that
islands coalesce to form a uniform layer, $\Delta/\lambda\to\infty$,
corresponding to a peristaltic regime $U_x(x,t)=0$.

Arguments similar to those leading to Eqs. (\ref{modification3}) and (\ref{zeta sigma}) 
can be used to estimate the stress perturbation under an island. 
The island's structure can be likened to that of a lamellar armor, in which
small metallic plates are laced into rows allowing
flexibility in the normal direction. Of course this flexibility is lost when the length scale
of the deformation (i.e. $\lambda$) is of the same order of the size of the lamellae (i.e.
the disks, $R$).

For the tangential stress, the role of $R$ is replaced by $\Delta$:
\beq
\langle\tau_{xz}\rangle=\frac{11\mu\alpha\Delta^2}{64}\partial_x^2 U_x
\tanh(\hat\alpha\psi).
\label{stress=1}
\eeq
From Eq. (\ref{gamma}), the expression for $\langle\tau_{xz}\rangle$ in Eq. (\ref{stress=1}) 
is $O(\epsilon_k^{-2})$ larger than the one in the first of Eq. 
(\ref{modification3}).

If we assume that the disks remain free 
to move vertically as in the dilute case, the normal stress will continue to be
generated by the resistance of the disks to bending. Thus, even if  
$\langle\pi_{zz}\rangle$ is going to be modified with respect to Eq. (\ref{modification3}),
its magnitude will be fixed by $R$ and will go to zero in the 
$\epsilon_k,\delta/R\to 0$ limit of
a continuous, horizontally homogeneous, but immaterial surface layer. This means that
to leading order in $\epsilon_k$ the normal stress at the surface can be neglected.

It is interesting to note that in the present situation, the condition $\epsilon_\alpha\ll 1$
in Eq. (\ref{geometric constraint}) loses meaning and should be replaced by
$\lambda_\alpha\ll\Delta$, which, for $\Delta\approx\lambda$, is always going to be 
satisfied. 

We are now in the position 
to derive a dispersion relation in the close-packing regime, along the line of
the procedure which leads to Eq. (\ref{reldisp3}) in the dilute case.
To allow comparison, at least in principle, with data from wave tank experiments, we allow 
$\hat H<\infty$. 

Equation (\ref{eq1}) for the tangential stress balance is then replaced by
\begin{eqnarray}
&&(\gamma_\psi\hat\nu^{1/2}\hat\alpha\hat k+2\hat\nu\hat k^2)\PhiU_+
+(\gamma_\psi\hat\nu^{1/2}\hat\alpha\hat k-2\hat\nu\hat k^2)\PhiU_-
\nonumber
\\
&&-(1+2\hat\nu\hat k^2)(\AU_++\AU_-)
-\gamma_\psi\ (1-\im\hat\nu\hat k^2)^{1/2}\ (\AU_+-\AU_-)=0,
\label{eq1bis}
\end{eqnarray} 
where $\gamma_\psi=\gamma\tanh\hat\alpha\psi$.
We note that the disks contribute to the dynamics at $O(\hat\nu^{1/2})$ while in the dilute
limit, they contribute only at $O(\hat\nu)$.

Considering a finite depth regime
implies that we must expand around $\hat k^\smalze=\hkH$, that is the solution to the
dispersion relation of gravity waves in a basin of depth $H$:
\beq
\hkH=\qH(\hkH).
\eeq
This means that we must expand in Eq. (\ref{eq4})
\beq
\qHh(\hat k)=\hkH+(\hkH-\hat k^\smalun\hat H)(\hkH^2-1)\psi\hat\nu^{1/2}+\ldots
\eeq

Putting to system Eq. (\ref{eq1bis}) 
with Eqs. (\ref{eq2}) to (\ref{eq4}), and expanding the resulting 
secular equation to $O(\hat\nu^{1/2})$ gives the dispersion relation
\beq
\frac{\hat k}{\hkH} \simeq 1+\hat\nu^{1/2}\left\{
\frac{\im\hat\alpha \gamma\hatrho\hkH^2\tanh(\hat\alpha\psi)}{(1+\gamma)[1+(\hkH^2-1)\hat H]}
+\frac{(1-\hatrho)(\hkH^2-1)\psi}
{1+(\hkH^2-1)\hat H}\right\},
\label{reldisp4}
\eeq
where the
normal stress from the disks, that is $O(\hat\nu)$, is disregarded.
The second term in braces in 
Eq. (\ref{reldisp4}) is the correction that would be produced by an inviscid layer
of density different from the rest of the column, which is just a mass-loading effect
(recall that $\psi\hat\nu^{1/2}=\hat h$, which is independent of $\nu$). The disks
are accounted for by the first term in braces. 

\section{Discussion}
\label{Discussion}
\subsection{The dilute theory}
The theory has been derived for small 
$f$, $\epsilon_\alpha$, $\epsilon_k$ and $\delta/R$.
The dispersion relation  Eq. (\ref{reldisp3}) accounts for the stresses by the viscous
layer and by the disks, but not for the disks mutual interaction. In most situations
involving pancake ice, such conditions are not fully satisfied. The dilute model can 
nevertheless be used in intermediate regimes, provided the pancake concentration
is not too high and the locking mechanism described in the previous section does not
set in. The situation as regards the other expansion parameters $\epsilon_k$ and 
$\epsilon_\alpha$ is not as dramatic and we expect that the theory is able to 
provide order of magnitude estimates also when the condition $\epsilon_{k,\alpha}\ll 1$
is not strictly satisfied.

An aspect that is worthwhile studying is the relative contribution 
by the viscous layer and by the disks to the wave dynamics.
The two relevant limits 
Eqs. (\ref{basic}) and (\ref{vis}) of the dispersion relation Eq. (\ref{reldisp3}) 
correspond to situations in which the stress
by the disks and by the viscous layer, respectively, are dominant. 

The magnitude of the
contribution to the dispersion relation Eq. (\ref{reldisp3})
from the stress by the viscous layer and from the tangential and 
normal stress components by the disks can be estimated as
\beq
O(\epsilon_\alpha^3\epsilon_k^3),\quad O(f\epsilon_\alpha\epsilon_k^3)\quad {\rm and}\quad
O(f\epsilon_k^4),
\label{estimates}
\eeq
respectively.  

Parameters typical of pancake ice in the ocean  are $\nu=0.01\,{\rm m^2/s},\ R<1\,{\rm m}
\ \kdeep=0.06\,{\rm m^{-1}}$), corresponding to $\epsilon_\alpha\approx 0.2$.  
To maximize the effect of the disks we
set nominally $f=1$ in the analysis that follows.

\begin{figure}
\begin{center}
\begin{minipage}{0.49\columnwidth}
\includegraphics[width=\columnwidth]{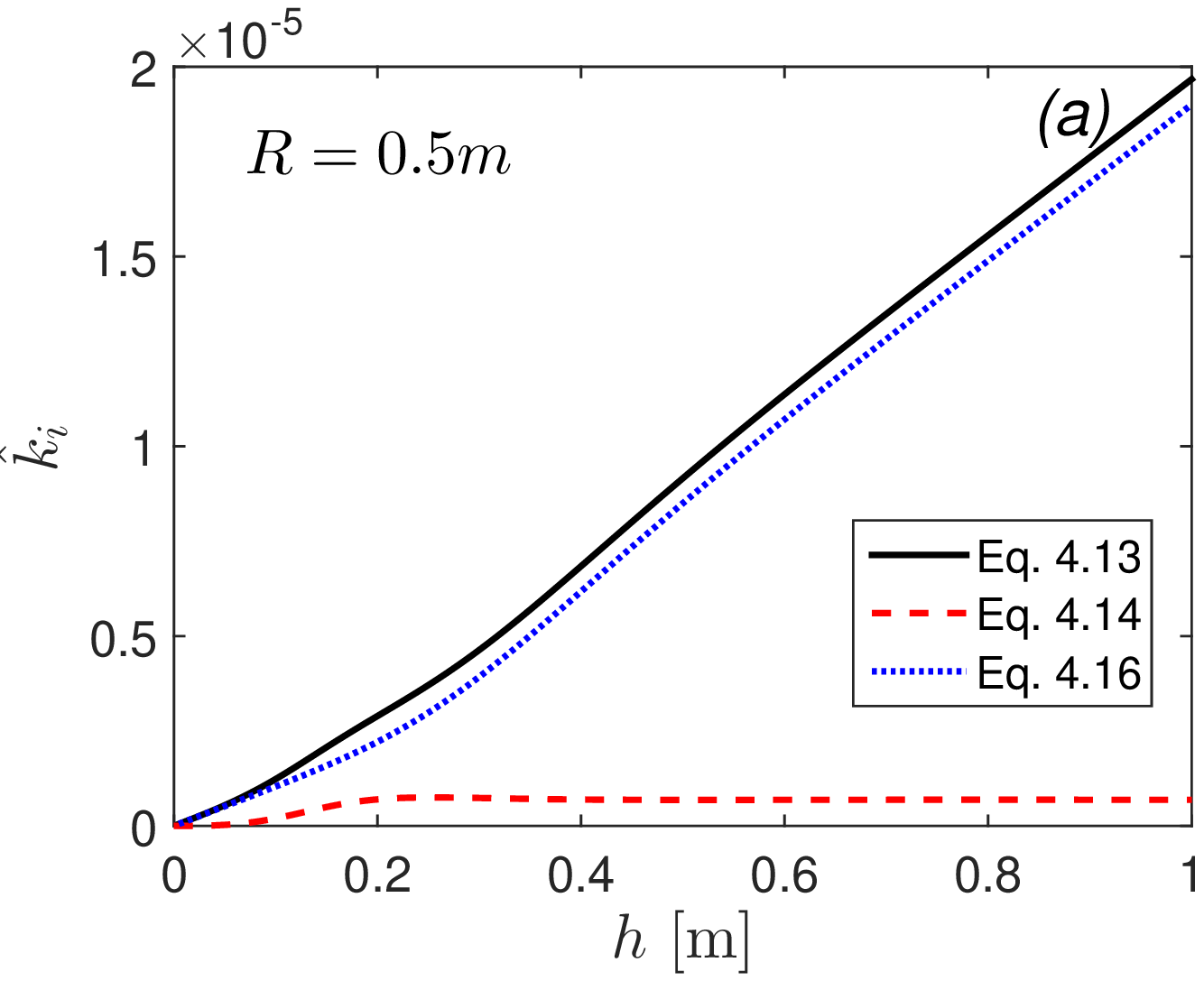}
\end{minipage}
\begin{minipage}{0.49\columnwidth}
\includegraphics[width=\columnwidth]{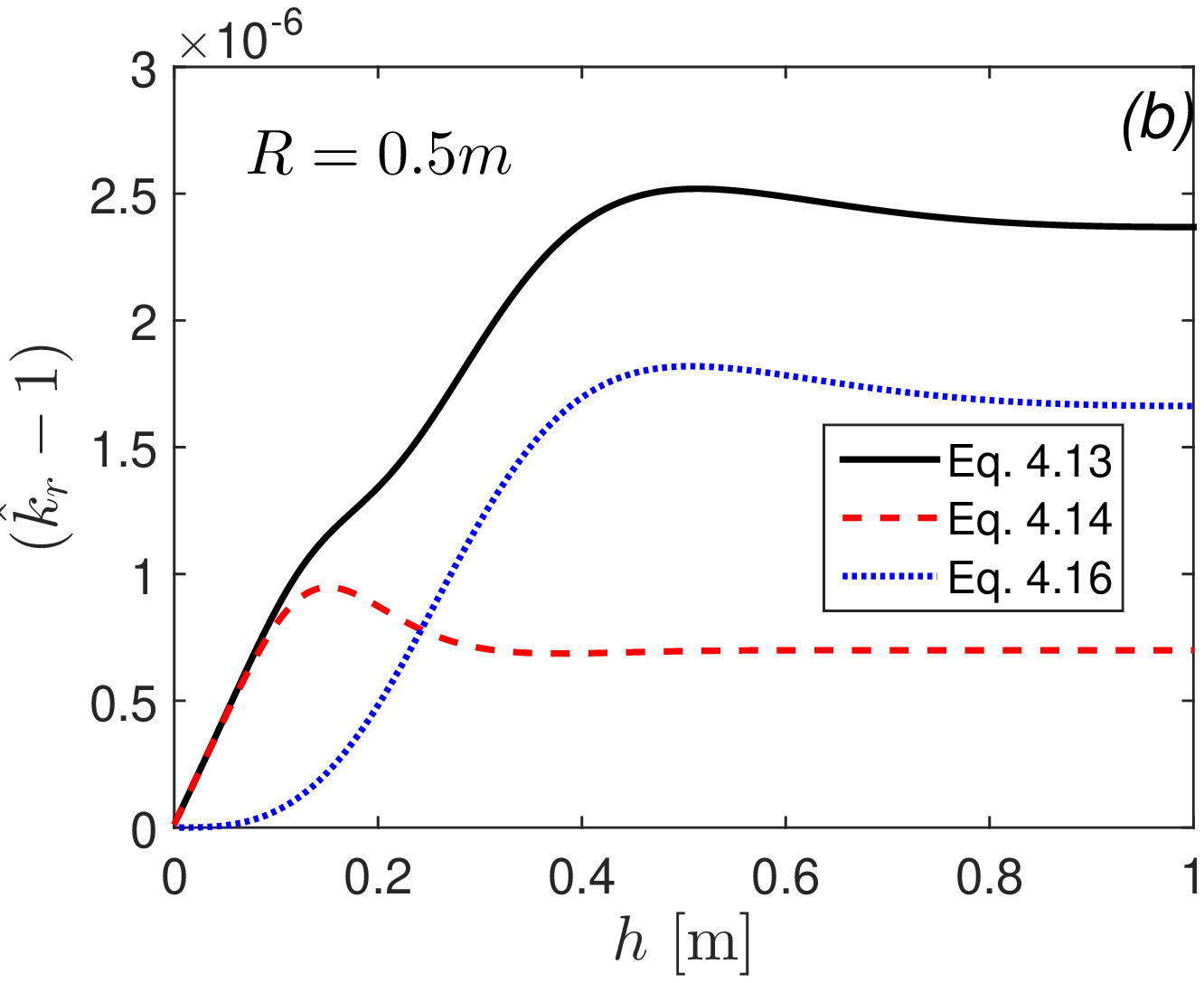}
\end{minipage}
\begin{minipage}{0.49\columnwidth}
\includegraphics[width=\columnwidth]{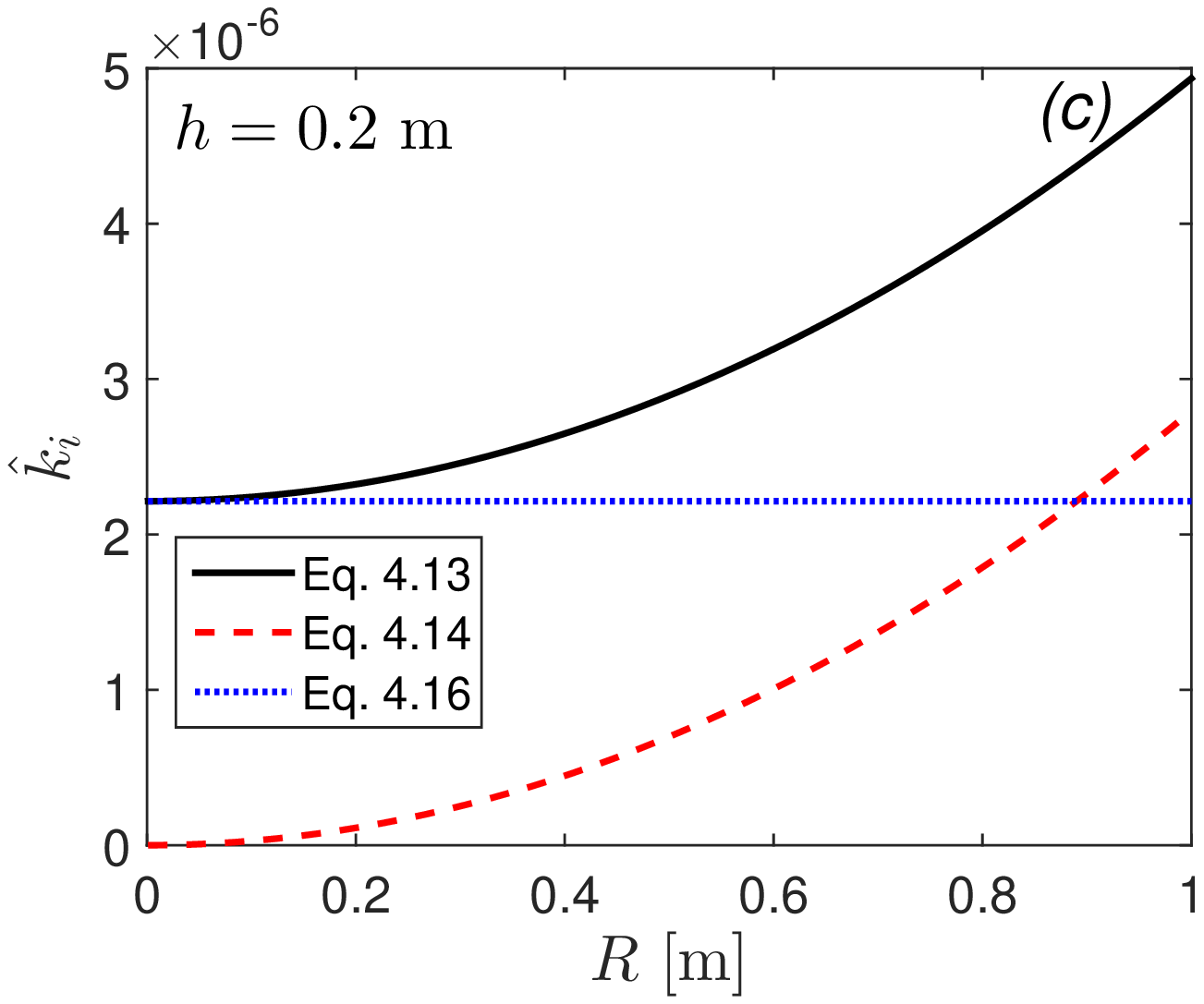}
\end{minipage}
\begin{minipage}{0.49\columnwidth}
\includegraphics[width=\columnwidth]{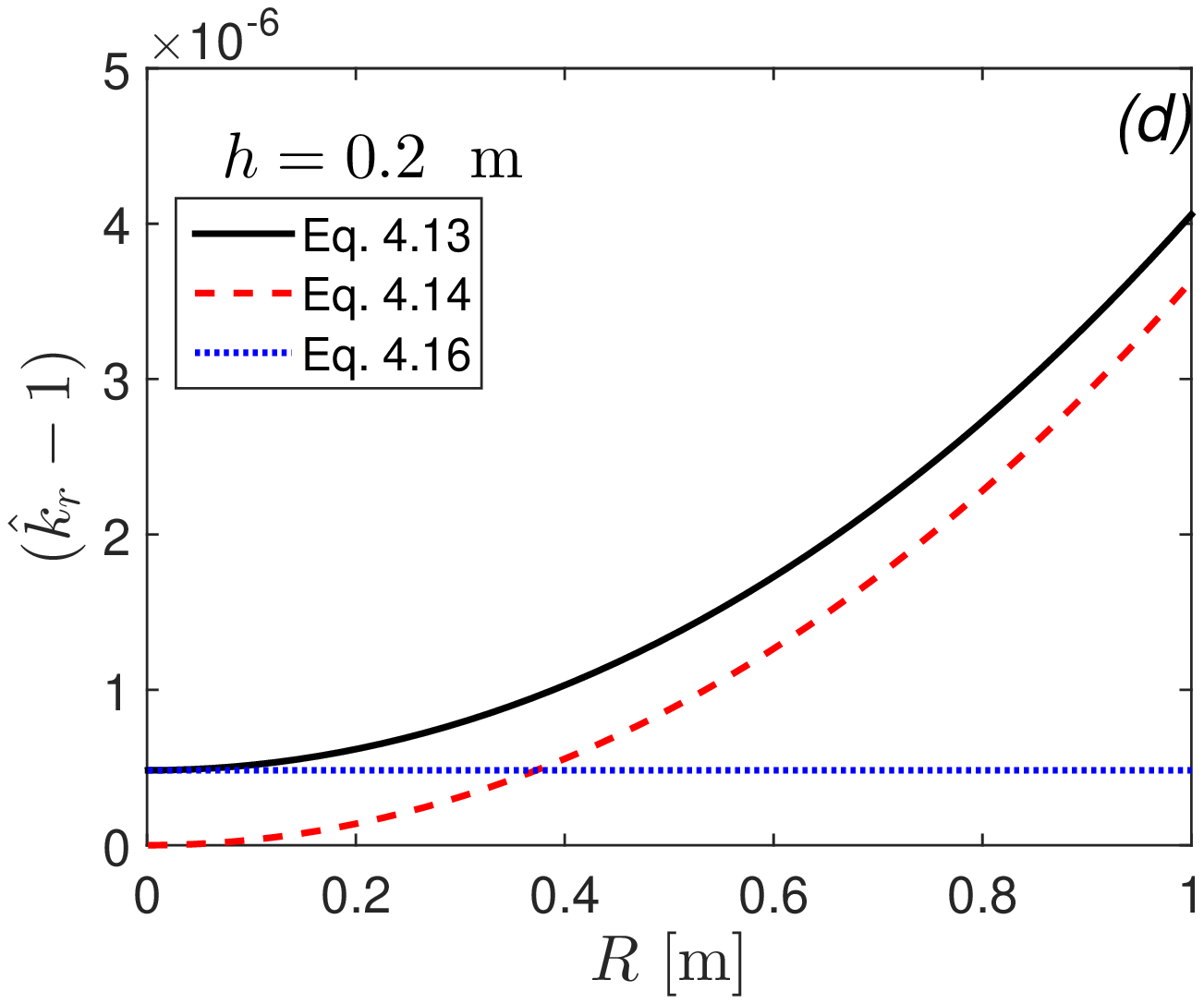}
\end{minipage}
\end{center}
\caption{Plots of wave damping (panels (\textit{a}) and (\textit{c})),
and wave dispersion (panels (\textit{b}) and (\textit{d})), in a 
typical pancake ice scenario. Values of the parameters:
$\nu=10^{-2}\,{\rm m^2/s}$, $\kdeep=0.06\,{\rm m^{-1}}$, $\hatrho=0.917$, 
corresponding to $\lambda_{\alpha}=0.1142\,$m and $\hat{\nu}=4.69\cdot
10^{-5}$'. The choice $R=0.5$ m in panels ($a$) and ($b$) leads to $\xi=0.13$.  
The choice $h=0.2$ m in in panels $(c)$ and ($d$) leads to 
$\psi=1.75$. In all panels $f=1$.
}
\label{eq_413_nu1e2}
\end{figure}

As shown in Fig. \ref{eq_413_nu1e2}, there are situations in which, even though $f=1$, 
the viscous layer dominates over the effect of the disks.
This corresponds to the $\xi\to 0$ limit of Eq. (\ref{reldisp3}), which is Eq. (\ref{vis}).
The effect is more pronounced for damping than for dispersion.  Inspection 
of Eq. (\ref{reldisp3}) and of its limit forms tells us that the viscous layer's effect is 
mainly damping of the waves, while the disks produce damping and dispersion that are
of the same order.
The damping by the viscous layer turns out to be larger than that by the disks 
($\epsilon_\alpha$ is not small
enough compared to the numerical coefficients in Eq. (\ref{reldisp3})). This implies that
damping dominates over dispersion in most of the parameter range considered, and
that the effect of the disks on wave damping is small.  The effect 
decreases at larger $h$ and smaller $R$.

The situation is different as regards wave dispersion, as the viscous layer 
contribution to $\hat k_r-1$ is much smaller than to $\hat k_i$. This has the consequence
that when the viscous layer is 
thin enough (less than $\approx\lambda_\alpha$ in thickness), 
the disks dominate dispersion.

\begin{figure}
\begin{center}
\begin{minipage}{0.49\columnwidth}
\includegraphics[width=\columnwidth]{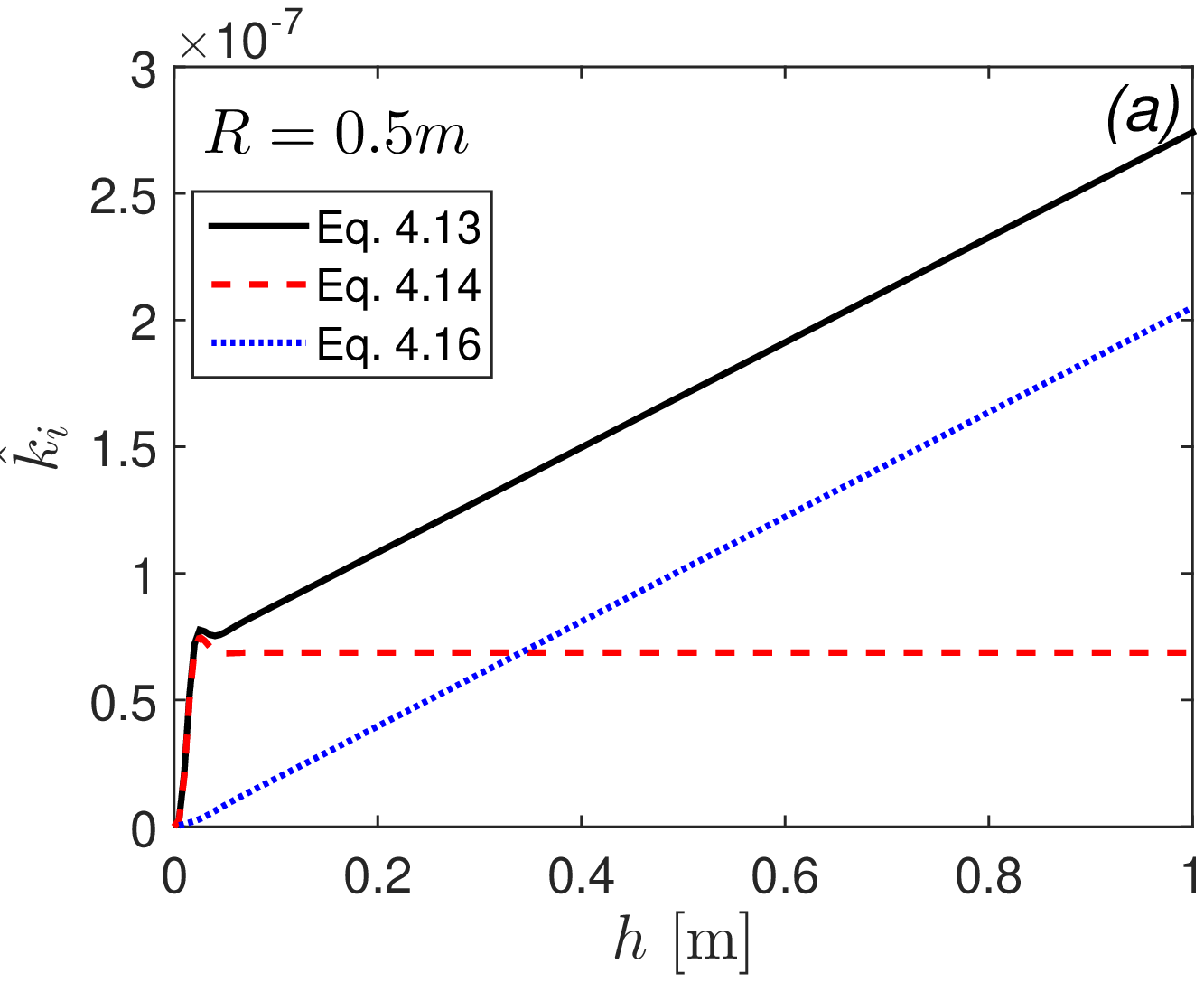}
\end{minipage}
\begin{minipage}{0.49\columnwidth}
\includegraphics[width=\columnwidth]{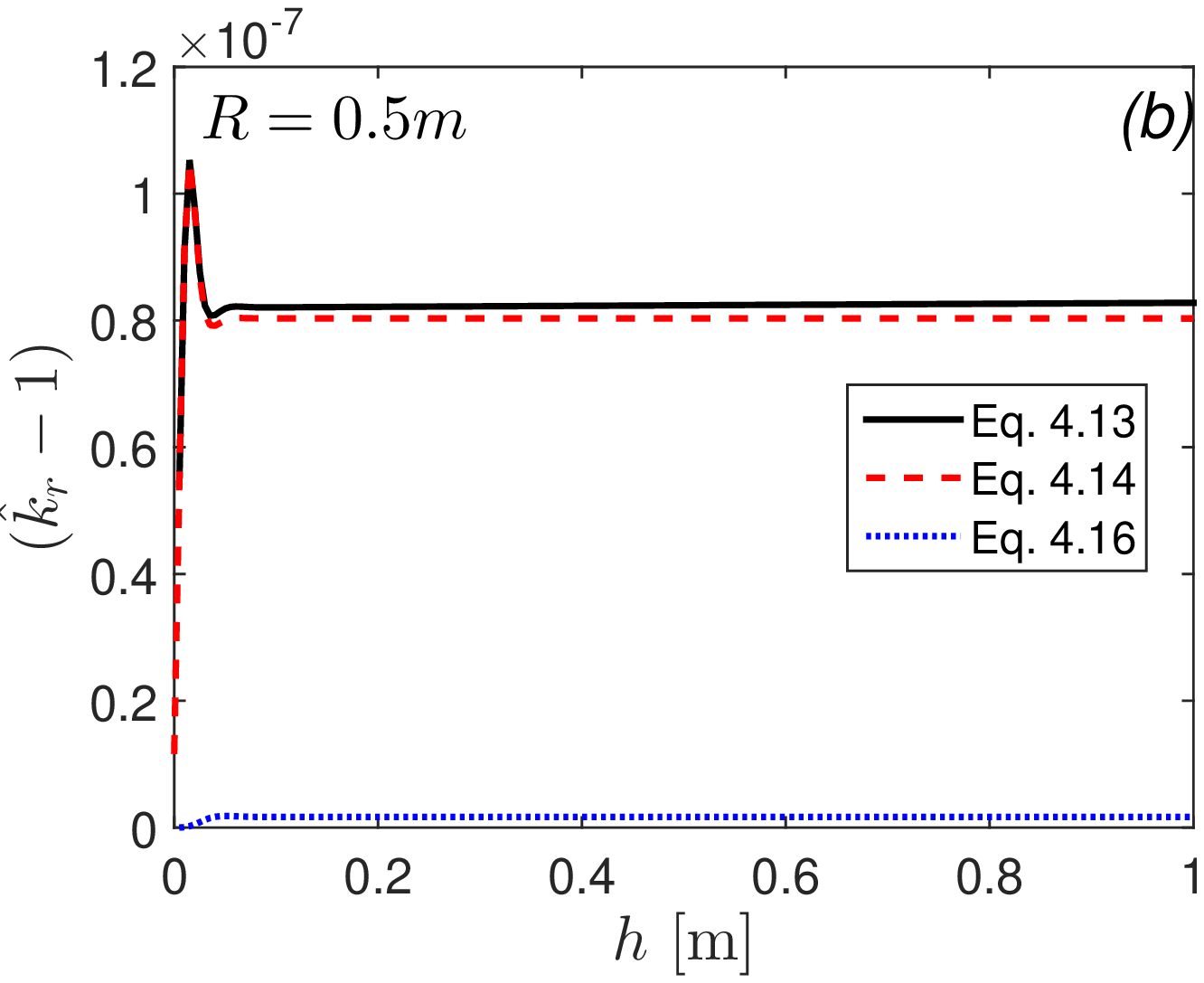}
\end{minipage}
\begin{minipage}{0.49\columnwidth}
\includegraphics[width=\columnwidth]{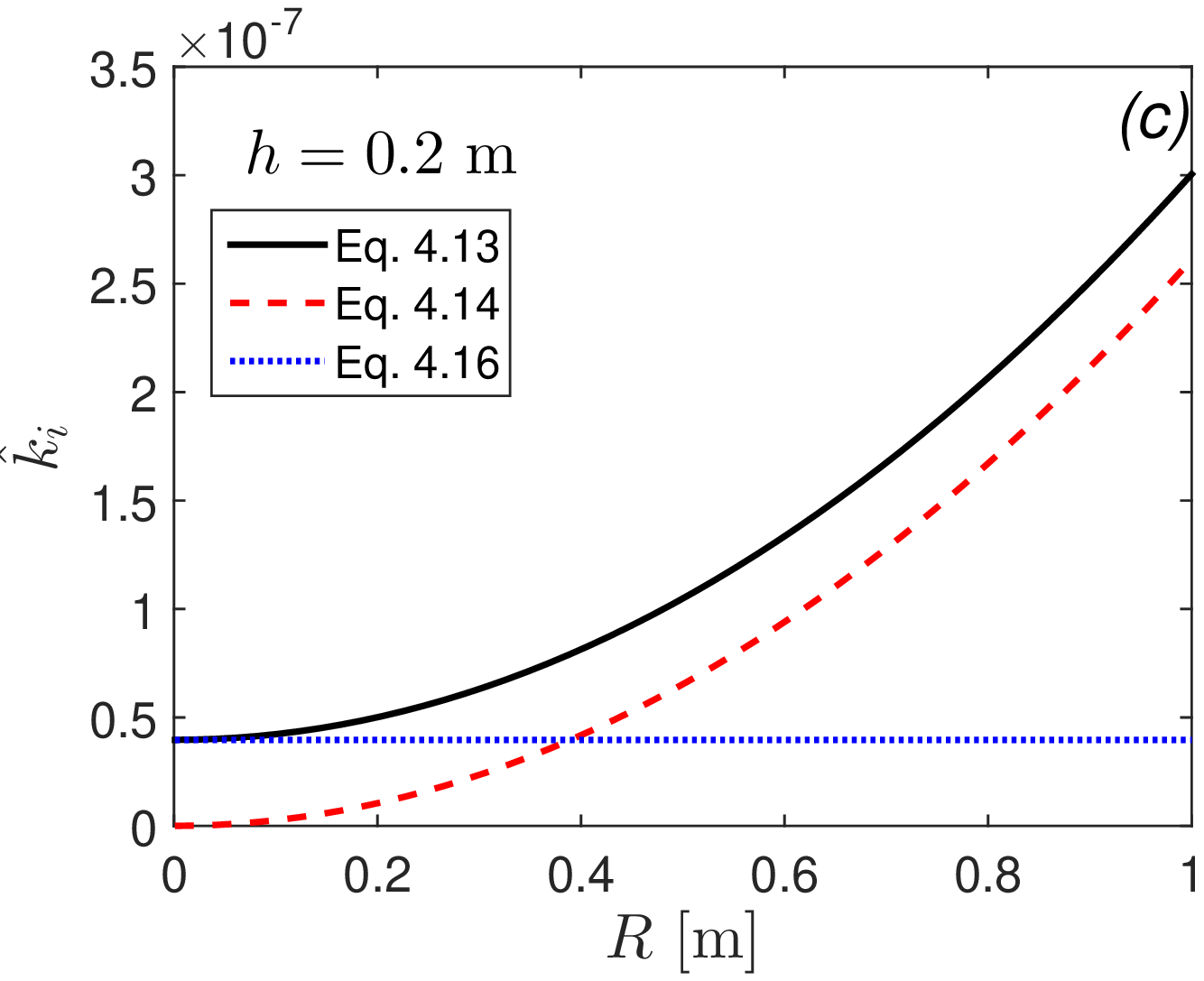}
\end{minipage}
\begin{minipage}{0.49\columnwidth}
\includegraphics[width=\columnwidth]{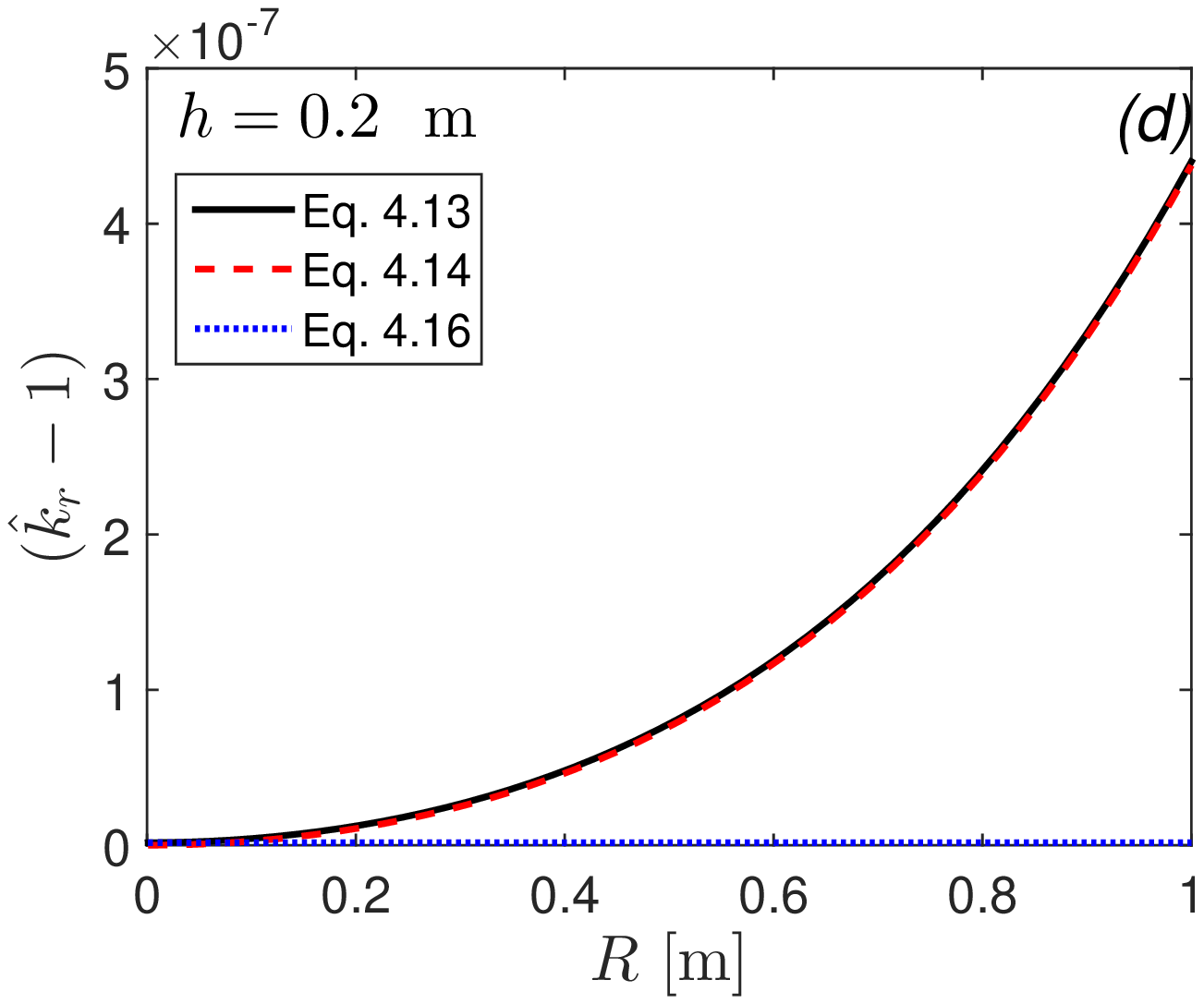}
\end{minipage}
\end{center}
\caption{Plots of wave damping (panels (\textit{a}) and (\textit{c})),
and wave dispersion (panels (\textit{b}) and (\textit{d})), in a
possible pancake ice - mixed oil scenario. Effective viscosity
$\nu=10^{-4}\,{\rm m^2/s}$; values of $\kdeep$, $\hatrho$ and $f$ as 
in Fig. \ref{eq_413_nu1e2}.
}
\label{eq_413_nu1e4}
\end{figure}

The dispersion relation is drastically modified 
when $\epsilon_k$ and $\epsilon_\alpha$ are both small
and $\xi=\epsilon_k/\epsilon_\alpha=O(1)$. 
Such a condition could be realized, in the range of $h$ and $R$ of Fig.  \ref{eq_413_nu1e2},
using a smaller viscosity.  
In the context of pancake ice, examples in which a smaller viscosity could be 
considered are generally related to the presence of oil: situations include
oil spilling under pancake ice and oil incorporated into the ice as grease 
ice is formed \citep{fingas03}. We repeat in Fig. \ref{eq_413_nu1e4} the analysis
in Fig.  \ref{eq_413_nu1e2} adopting $\nu=10^{-4}\,{\rm m^2/s}$.
In this case, we see that wave dispersion is dominated by the disks, while 
damping is dominated by the disks only for small $h$ and large $R$. In this parameter range, 
damping and dispersion are of the same order of magnitude. 
The slow dependence of the disk stress on $h$ 
is due to the fact that for small $\nu$,
$\lambda_\alpha$ is small as well, $\psi$ is consequently large, and the tangential stress in 
Eqs. (\ref{modification3}) and (\ref{zeta sigma}) reaches a plateau.

It is to be noted that the wave damping does not grow without bound for
$\nu\to\infty$, instead, it first reaches a maximum for $h\sim\lambda_\alpha$, and then 
goes to zero for $\nu\to\infty$ (the first term in braces to RHS of Eq. (\ref{reldisp4})
becomes purely real in the limit, with $\im\hat\alpha\tanh(\hat\alpha\psi)\simeq-\psi$).
The limit corresponds to the viscous layer behaving as a rigid lid with the shape of the wave,
which is transported by the wave itself.
From analysis of Eq. (\ref{vis}), we see that the same limit cannot be achieved---within 
perturbation theory at least---in the case of a simple viscous layer.

\subsection{The close-packing model}
The model has been introduced to take into account the reduced relative mobility of the
disks for $f\approx 1$. The dispersion relation 
Eq. (\ref{reldisp4}), which is valid in the limit 
\beq
R\ll\Delta,\lambda,
\label{R Delta lambda}
\eeq
describes a situation in which both wave damping and dispersion are greatly increased
with respect to the prediction of the dilute theory, for comparable values of the 
surface fraction $f$.

\begin{figure}
\begin{center}
\begin{minipage}{0.49\columnwidth}
\includegraphics[width=\columnwidth]{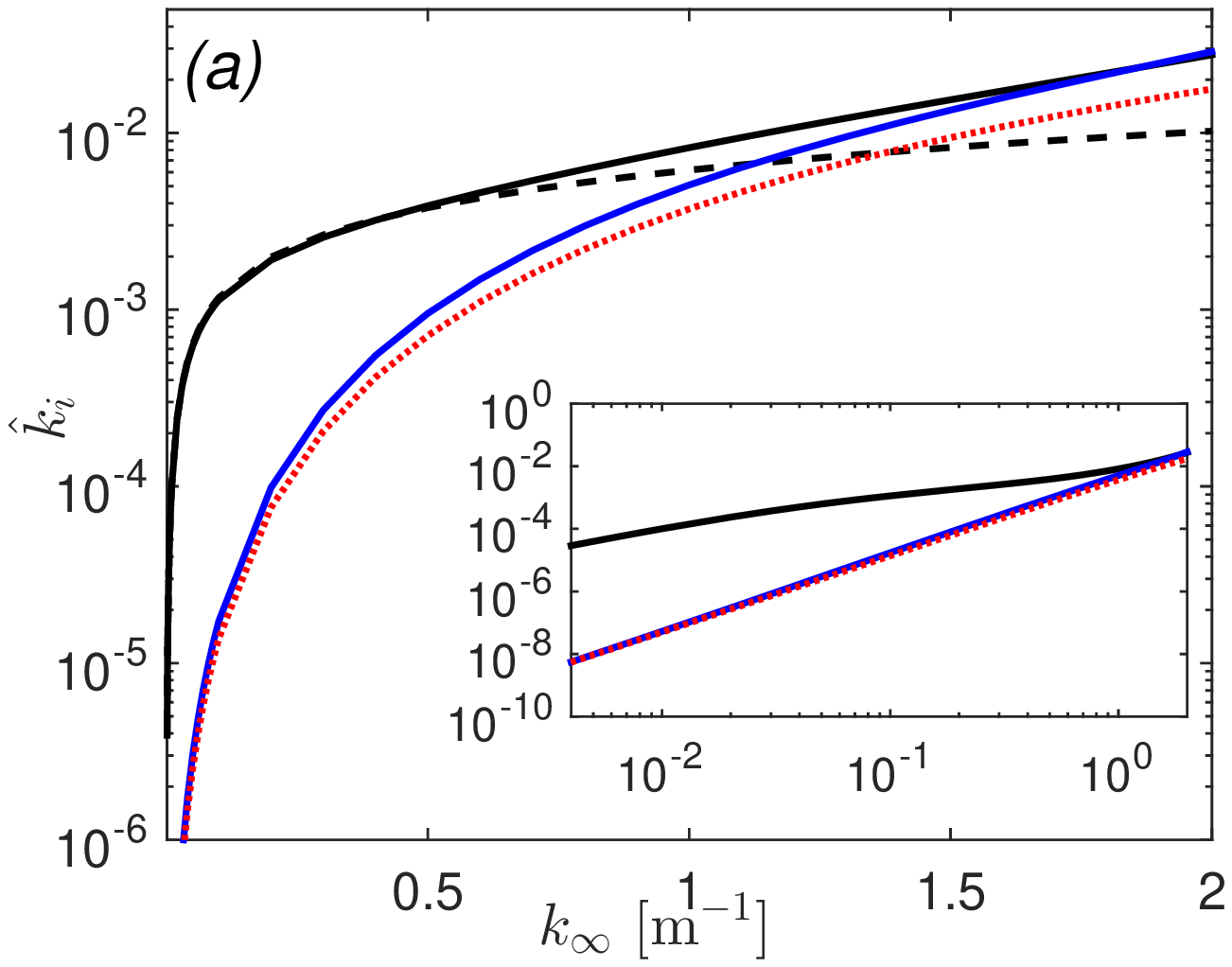}
\end{minipage}
\begin{minipage}{0.49\columnwidth}
\includegraphics[width=\columnwidth]{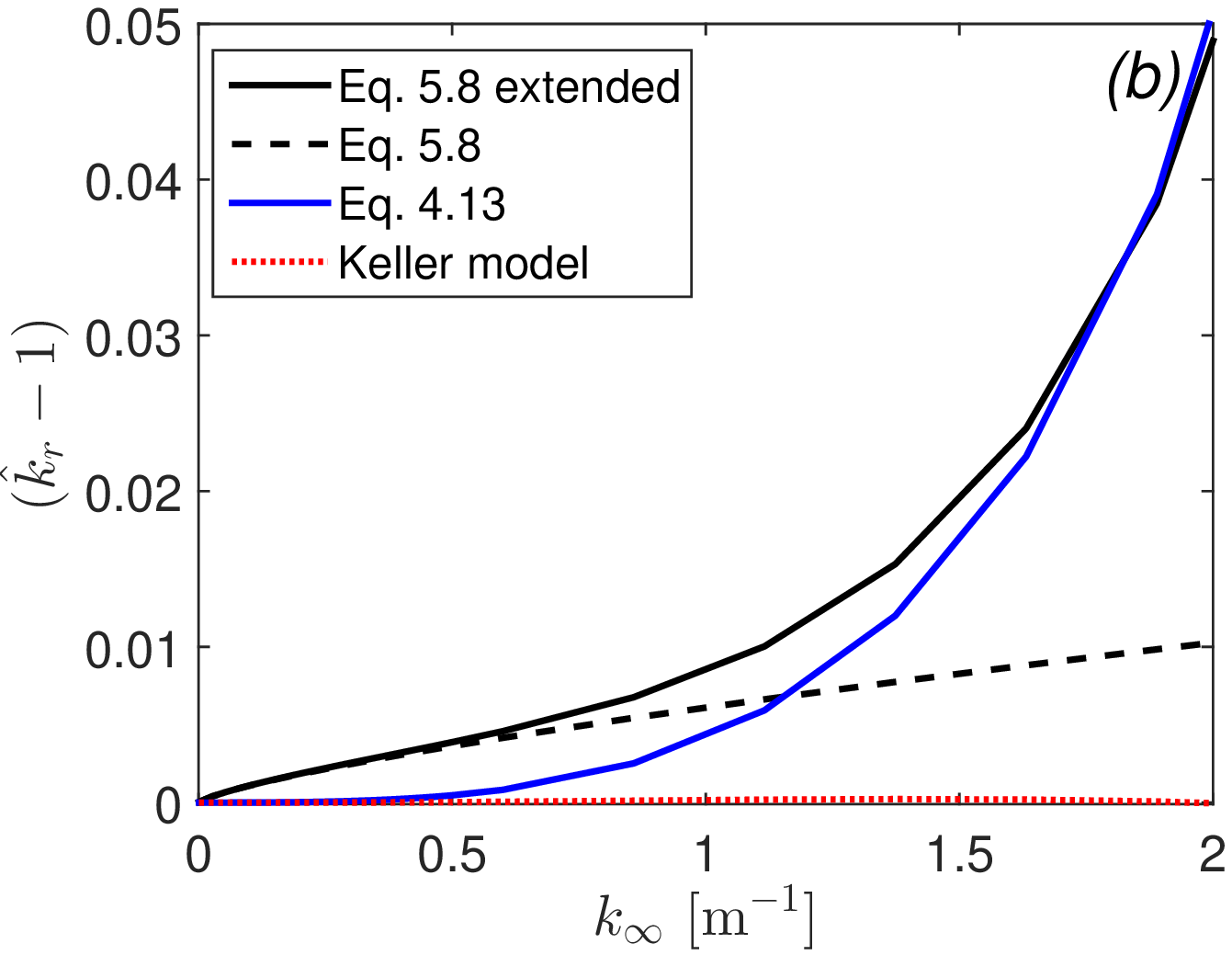}
\end{minipage}
\caption{Comparison of the the predictions by the close-packing model and its extended version, 
the dilute theory and 
the Keller theory, in a typical pancake ice scenario. Values of the parameters:
$\nu=10^{-2}\,{\rm m^2/s}$, $\hatrho=0.917$, $h=0.2\,$m, $R=0.5\,$m, $f=1$ and $\gamma=0.2$.
Panel (\textit{a}) wave damping; panel (\textit{b}) dispersion.
}
\label{short}
\end{center}
\end{figure}

The dispersion relation Eq. (\ref{reldisp4}) neglects normal stress contributions.
Such contributions can be taken into account by 
an extended version of the model in which Eqs.
(\ref{eq1bis}) and (\ref{eq2}) to (\ref{eq4}) are solved without 
approximations.\footnote{It should be stressed
that the extended version of Eq. (\ref{reldisp4}) can provide
only an estimate of the behaviour of the dispersion relation at small wavelengths, since
only a part of the higher order contributions in $\epsilon_k$ in the perturbation expansion
for the stresses (Eqs. (\ref{modification3}) and (\ref{zeta sigma}))
is taken into account.}
This reveals  that the simplified model works well as long as
$\Delta/R\gtrsim 4$ (see Fig. \ref{short}).
For $\Delta\sim R$, the ``extended'' close-packing model
merges with the dilute theory Eq. (\ref{reldisp3}). As regards wave damping, 
we see in Fig. \ref{short}{\it a} 
that for a typical pancake ice scenario, 
with $\nu=0.01\ {\rm m^2/s}$, $R=0.5$ m and fixed $\gamma=O(1)$,
this merging occurs for very short waves.

As regards wave dispersion, Fig. \ref{short}{\it b} illustrates that both Eqs. (\ref{reldisp3})
and (\ref{reldisp4}), 
and the extension of the second to $\epsilon_k\approx 1$, lead to a decrease of the
phase velocity. As in the case of damping, the close-packing model gives a result that is
orders of magnitude larger than that of the dilute theory at large wavelengths. We have included
for reference the prediction by the Keller theory \citep{keller98}, which,
for all values of $\kdeep$, disappears in front of the contribution from the
pancakes in close packing conditions.

We can compare the results of the close-packing model with those 
of a viscoelastic model such as the one by \cite{wang10}. 
The composite pancake-grease-ice layer is treated as a
homogeneous Voigt medium with complex viscosity
\beq
\nu_e=\nu+\frac{\im G}{\varrho\omega}.
\label{Voigt}
\eeq
We take for the elastic modulus
$G=10^3$ Pa (a value in the range of gels, but still much less 
than what would be observed in solid ice), and keep considering a shear viscosity typical of
grease ice, $\nu=10^{-2}\ {\rm m^2/s}$.
This choice guarantees that, in the small
to moderate $\kdeep$ range considered, $\hat\nu_e=\kdeep^{3/2}g^{-1/2}\nu_e$ is small.
The full dispersion relation, Eq. (45) in \cite{wang10}, can then
be approximated with our Eq. (\ref{vis})
by setting $H\to\infty$ and substituting $\hat\nu\to\hat\nu_e$.
\begin{figure}
\begin{center}
\begin{minipage}{0.49\columnwidth}
\includegraphics[width=\columnwidth]{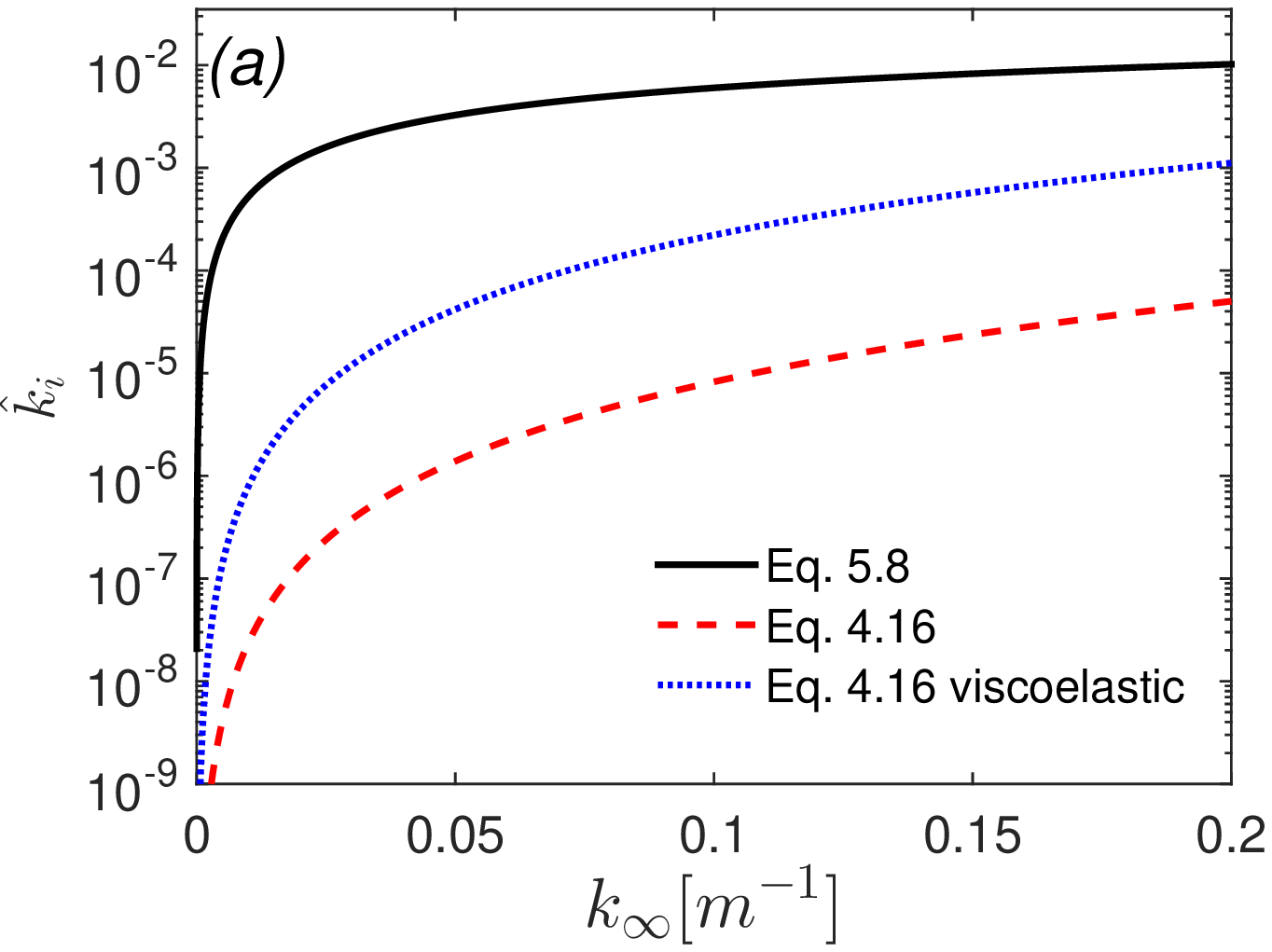}
\end{minipage}
\begin{minipage}{0.49\columnwidth}
\includegraphics[width=\columnwidth]{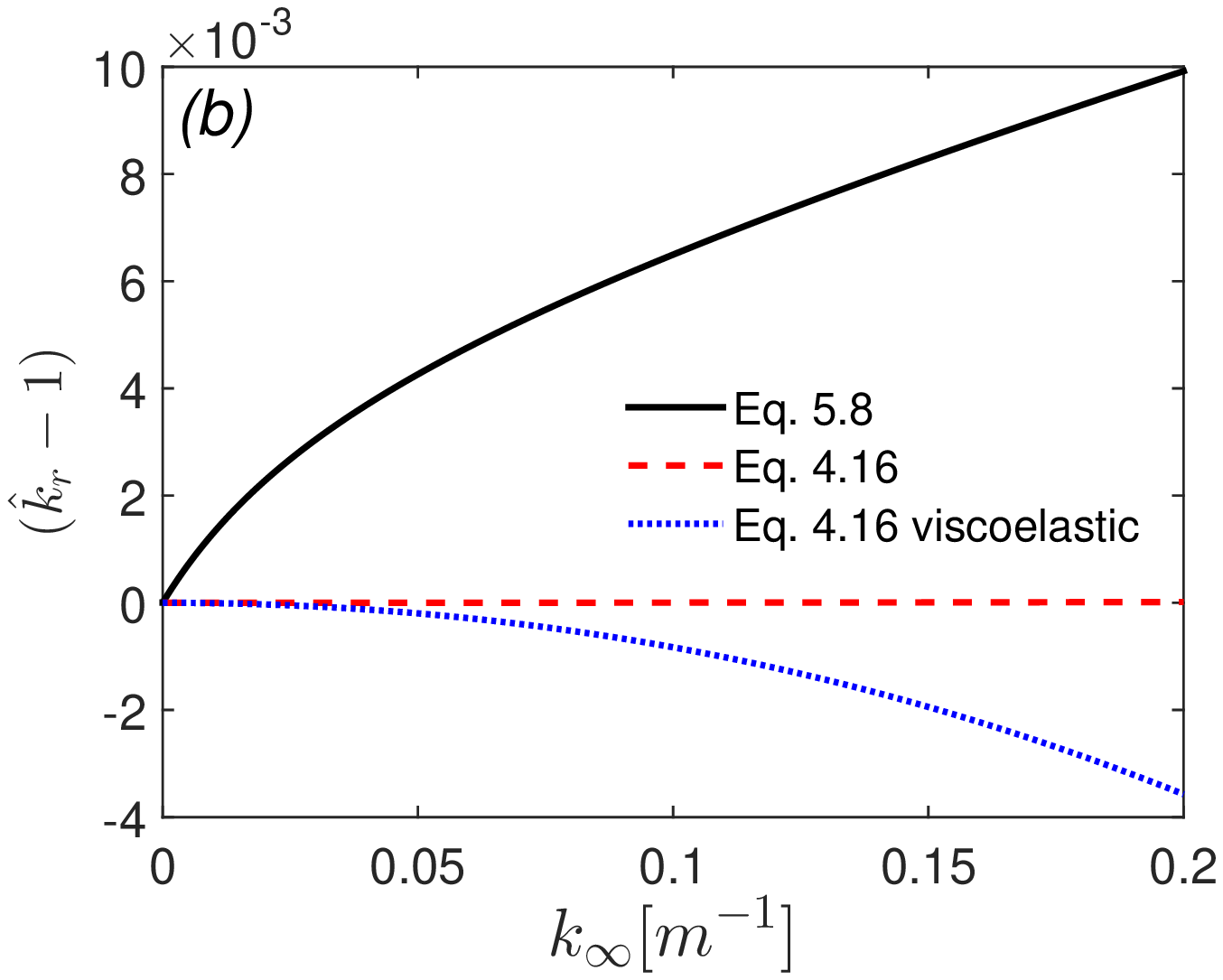}
\end{minipage}
\caption{Comparison of the the predictions by the close-packing model, 
the viscous layer model and the viscoelastic layer model. Values of the parameters:
$\nu=10^{-2}\,{\rm m^2/s}$, $\hatrho=0.917$, $h=0.2\,$m, $R=0.5\,$m, $\gamma=6$, $G=10^3\,{\rm Pa}$.
Panel (\textit{a}) wave damping; panel (\textit{b}) dispersion.
}
\label{viscoel}
\end{center}
\end{figure}

As shown in Fig. \ref{viscoel},
the viscoelastic model
predicts a wave damping much smaller than the close-packing model, Eq. (\ref{reldisp4}).
Even smaller values of $\hat k_i$ are predicted in the purely viscous model, that is
Eq. (\ref{vis}) in its original form with $\nu=0.01\ {\rm m^2/s}$ and $G=0$.
The value of $\gamma$ adopted is close to the inextensible membrane limit; however, different
choices do not produce dramatic modifications (see Fig. \ref{short}).
The viscoelastic and close-packing model lead to values 
of the dispersion modification 
$\hat k_r-1$ of the same order of magnitude, although with opposite sign.
Comparison with data from synthetic aperture radar imaginery by \cite{wadhams91} (see Table
1 in that reference),
suggest that $\hat k_r-1>0$ as in the close-packing model, even though it must
be mentioned that such data referred to an inhomogeneous situation, in which regions 
with just grease ice and regions with grease and pancake ice were both present.

We have checked the validity of the close-packing model against some real 
field data on sea ice.
We have considered the Bering Sea data of 7th February 1983  on wave damping 
reported by  \cite{wadhams88}. The data refer to waves propagating
in ocean covered with ice floes of radius $R\approx 10$ m. 

A fit of the data by Eq. (\ref{reldisp4}) has been obtained for the reference value
of the viscosity in the top layer $\nu=0.01\ {\rm m^2/s}$. 
The top layer viscosity may be due to the presence of grease ice, but in principle
an eddy viscosity contribution may also be present due to the underwater stresses generated
by the wind.

Best fit of Eq. (\ref{reldisp4}) by least squares
gives $h=0.14$ m, while $\gamma$ can take any value $\gtrsim 10$, corresponding 
to a peristaltic regime. Least square fit by the Keller model \citep{keller98} varying $\nu$ 
and $h$, gives $\nu=10\ {\rm m^2/s}$ and $h=0.28$ m. 
In order for such a model 
to generate a wave damping
of the same order of magnitude, a much larger value
of the viscosity in the top layer must be adopted,
which seems rather unrealistic. 

Both models fail to predict the apparent rollover in 
the spectrum at $\kdeep>0.06\ {\rm m^{-1}}$.
\begin{figure}
\begin{center}
\begin{minipage}{0.49\columnwidth}
\includegraphics[width=\columnwidth]{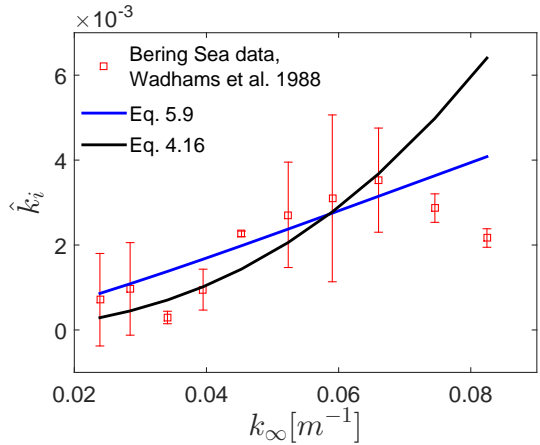}
\end{minipage}
\caption{Comparison of field data on wave damping by ice floes ($R=10\,$m)
\citep{wadhams88} and
prediction by the Keller model and close-packing model Eqs. (\ref{vis}) and (\ref{reldisp4}).
Values of the parameters: 
$\nu=0.01\,{\rm m^2/s}$, $h=0.14\,$m, $\gamma\to\infty$
(close-packing model); $\nu=10\,{\rm m^2/s}$, $h=0.21\,$m (Keller model).
}
\label{floes}
\end{center}
\end{figure}
%
%
%
%

%

\section{Conclusion}
\label{Conclusion}
We have studied the propagation of gravity waves in a water body covered by
a distribution of thin disks embedded in a viscous layer.
We have described the wave dynamics as a function of the
surface fraction of the disks $f$, and of
the relevant scales of the problem: the disk radius $R$; the 
wavelength $\lambda$; the depth $h$ of the viscous layer; the thickness $\lambda_\alpha$
of the viscous boundary layer at the surface (see
Eq. (\ref{lambda_alpha})). 

We have provided an analytical theory valid in the limit $\lambda_\alpha\ll R\ll\lambda$,
$f\ll 1$. In such dilute limit, the interaction among disks is disregarded.
In the range $\lambda_\alpha\lesssim h$,
the role of control parameter is played by the quantity
$\xi\sim R^2/(\lambda\lambda_\alpha)$ defined in Eq. (\ref{dimensionless}).
In particular, the ratio of  
the normal and tangential stresses by the disks,
and the ratio of the contribution from the disks and from the viscous layer 
to wave damping, are both proportional to $\xi$. It is interesting to note
that, for values of the parameters compatible with pancake ice in the ocean (for which
the dilute theory, however, would not work),
the contribution to wave damping from the viscous layer would exceed that from
the disks.

We have used the dilute theory as a groundwork for the development of a macroscopic
model valid in a close-packing regime $f\approx 1$.

While in the dilute case, the surface stress is associated with surface strain rate on 
the scale of the individual disks, in the close-packing case it is the
whole disk layer that resists horizontal compression. The result is a dramatic increase
of the friction forces by the disks on the viscous layer, 
with wave damping and dispersion corrections
larger by orders of magnitude than predicted by the dilute theory.

An interesting point in the close-packing model
is the appearance of a new characteristic length $\Delta$,
which represents the extension of the regions in which disks 
are so closely packed to form a horizontally rigid structure,
and where tangential stress is maximum.

We have used the close-packing model to fit 
field data of wave propagation in ocean
covered with ice floes \citep{wadhams88,wadhams91}.  
We have compared the performance of the model with that of
the theory by \cite{keller98}.
Using values of the parameters compatible with presence of a grease ice layer or 
possibly a turbulent boundary layer 
(effective viscosity $\nu=0.01\,{\rm m^2/s}$),
we have observed that the data on wave damping reported by \cite{wadhams88}
can be fitted reasonably well with the close-packing 
model. In comparison, the Keller theory
would require much larger (and difficult to justify)
values of the effective viscosity. 

As far as wave dispersion is concerned, the close-packing model
predicts a decrease of phase velocity, which seems to agree with the data by
\cite{wadhams91}.

The close packing model fails to account for any rollover effect,
which are predicted instead by nonlinear models such as the one by \cite{shen98}. In principle,
nonlinear effects could be made to sneak in the close-packing model, by
taking seriously the interpretation of $\Delta$ as the extension of the
high compression regions of the wave field (the parameter $\gamma$ in Eq. (\ref{gamma}) would become
a function of $\U$). 
It is to be mentioned that rollover effects typically take place when the size of the disks
is comparable to the wavelength, in which case our theory ceases to be meaningful.
This prevents comparison with wave-tank data such as the ones in \cite{wang10a}.
Similar limitations exist also with field data on wave damping 
in ice covered ocean, when large floes are present \citep{kohout14}.
\\
\\
\noindent\textbf{Acknowledgements} 
We wish to thank Peter Wadhams for suggestions in the initial phase of
the work and Siddhanta Sabyasachi for helpfuld discussion. 
This research was supported by FP7 EU project ICE-ARC
(Grant agreement No. 603887) and by MIUR-PNRA, PANACEA project (Grant No. 2013/AN2.02).




\appendix
\section{Potential representation of time-dependent Stokes flows}
\label{Potential representation}
We decompose the fluid velocity as
\beq
\u=-\nabla\Phi+\nabla\times\A.
\eeq
Assuming incompressibility, we have that $\Phi$ is potential
\beq
\nabla\cdot\u=0\Rightarrow\nabla^2\Phi=0.
\eeq
The time-dependent Stokes equation is
\beq
\partial_t\u+\frac{1}{\varrho}\nabla P=\nu\nabla^2\u+\frac{1}{\varrho}\f,
\eeq
from which we get the vorticity equation
\beq
\partial_t[\nabla\times\u]=\nu\nabla^2[\nabla\times\u]
\eeq
(we consider for simplicity the case in which the force field is
of gradient type); in terms of potentials:
\beq
\partial_t[\nabla\nabla\cdot\A-\nabla^2\A]=
\nu\nabla^2[\nabla\nabla\cdot\A-\nabla^2\A],
\eeq
and, if we assume $\A$ divergenceless,
\beq
\partial_t\nabla^2\A=\nu\nabla^2\nabla^2\A.
\label{biharmonic}
\eeq
Equation (\ref{biharmonic})
has general solution $\A=\bar\A+\A'$, where
\beq
\partial_t\bar\A=\nu\nabla^2\bar\A,
\qquad
\nabla^2\A'=0.
\label{harmonic}
\eeq
If we continue to assume that $\A$ is divergenceless, 
we see that $\A'$ does not contribute to vorticity
\beq
\nabla\times[\nabla\times\A']=-\nabla^2\A'=0.
\eeq
We could decompose
\beq
\A'=\nabla q+\nabla\times\C,
\qquad\nabla^2 q=0.
\eeq
We have
\beq
\nabla^2\A'=\nabla\times\nabla^2\C=0\Rightarrow \nabla^2\C=
\nabla g,
\eeq
which allows us to write the contribution of $\A'$ to the velocity in the form
\beq
\nabla\times\A'=\nabla(\nabla\cdot\C-g).
\eeq
We see that adding a potential term $\A'$ to the vector potential
has the same effect as renormalizing the scalar potential:
\beq
\Phi\to\Phi+g-\nabla\cdot\C.
\eeq
In general, the equation for the scalar potential will be
\beq
-\partial_t\nabla(\Phi+g-\nabla\cdot\C)+\frac{1}{\varrho}\nabla P=\frac{1}{\varrho}\f.
\eeq
The equation will simplify if $\A'=0$, i.e. if we assume that
$\A$ obeys the first of Eq. (\ref{harmonic}). In this case we shall have,
taking $\f=-\nabla V$:
\beq
\partial_t\Phi=\frac{P+V}{\varrho}
\quad{\rm and}\quad
\partial_t\A=\nu\nabla^2\A,
\eeq
that are Eqs. (\ref{potential pressure}) and  (\ref{potential-equations}).

\section{Green function of the potential component}
\label{Green}
It is convenient to expand the Neumann Green function, Eq. (\ref{Neumann}), in angular harmonics:
\beq
G^N(\r,\rho_0)
=\frac{2}{\sqrt{\rho^2+\rho_0^2}}\sum_{m=-\infty}^{+\infty}
g_m^N\Big(\frac{r\rho_0}{\rho^2+\rho_0^2}\Big)\ex^{\im m(\phi_0-\phi_0')},
\eeq
with
\beq
g^N_m(x)=
\frac{1}{2\pi}\int_0^{2\pi}\d\phi
\frac{\ex^{-\im m\phi}}{\sqrt{1+x\cos\phi}}.
\eeq
Thus
\beq
\Phi^\smalze(\r)|_{z=0}=\sum_{m=-\infty}^{+\infty}\Phi^\smalze_m(\rho,0)\ex^{\im m\phi_0},
\eeq
where
\beq
\Phi^\smalze_m(\rho,0)=-\int_0^R\rho_0\,\d\rho_0\
\frac{u^\smalze_{m,z}(\rho_0,0)}{\sqrt{\rho^2+\rho_0^2}}
g^N_m\Big(\frac{\rho\rho_0}{\rho^2+\rho_0^2}\Big),
\label{gN}
\eeq
and similar expressions holding at higher orders. This allows to rewrite Eq. (\ref{papera4}) as
\begin{eqnarray}
\langle\pi_{xz}\rangle
&\simeq&\frac{f\mu\alpha}{2\pi R}\frac{\partial^2\bar U_x}{\partial\bar x^2}
\int_0^{2\pi}\d\phi\int_0^R\rho^2\,\d\rho\ \Big[\frac{\rho}{2R}(\cos^4\phi+\sin^22\phi)\Big]
\nonumber
\\
&-&\frac{f\mu\alpha}{4\pi}\frac{\partial^3\bar U_z}{\partial\bar x^3}
\int_0^{2\pi}\d\phi\int_0^R\rho^2\,\d\rho\ \Big\{\cos^2\phi
\nonumber
\\
&\times&\partial_\rho \int_0^R\rho_0\d\rho_0\  \frac{(\rho_0/R)^2-1}{\sqrt{\rho^2+\rho_0^2}}
g^N_0\Big(\frac{\rho\rho_0}{\rho^2+\rho_0^2}\Big)
+2\Big(\cos 2\phi\cos^2\phi\partial_\rho \nonumber
\\
&+&\frac{\sin^22\phi}{\rho}\Big)
\int_0^R\rho_0\,\d\rho_0\ \frac{(\rho_0/R)^2-1}{\sqrt{\rho^2+\rho_0^2}}
g^N_2\Big(\frac{\rho\rho_0}{\rho^2+\rho_0^2}\Big)\Big\}.
\end{eqnarray}
Carrying out the polar integrals and integrating by part in $R$ where necessary, we find
\begin{eqnarray}
\langle\pi_{xz}\rangle&\simeq&\frac{11f\mu R^2\alpha}{64}\frac{\partial^2\bar U_x}{\partial\bar x^2}
\nonumber
\\
&-&\frac{f\mu\alpha}{4}\frac{\partial^3\bar U_z}{\partial\bar x^3}
\int_0^R\rho^2\,\d\rho\ \Big\{\partial_\rho \int_0^R\rho_0\,\d\rho_0\ \frac{(\rho_0/R)^2-1}
{\sqrt{\rho^2+\rho_0^2}}
g^N_0\Big(\frac{\rho\rho_0}{\rho^2+\rho_0^2}\Big)
\nonumber
\\
&+&
\Big(\partial_\rho+\frac{2}{\rho}\Big)
\int_0^R\rho_0\d\rho_0\ \frac{(\rho_0/R)^2-1}{\sqrt{\rho^2+\rho_0^2}}
g^N_2\Big(\frac{\rho\rho_0}{\rho^2+\rho_0^2}\Big)\Big\}
\nonumber
\\
&=&
\frac{11f\mu R^2\alpha}{64}\frac{\partial^2\bar U_x}{\partial\bar x^2}
-\frac{Bf\mu\alpha R^3}{2}\frac{\partial^3\bar U_z}{\partial\bar x^3},
\end{eqnarray}
with
\beq
B=\int_0^1\rho\,\d\rho\ \int_0^1\rho_0\d\rho_0\ \frac{1-\rho_0^2}{\sqrt{\rho^2+\rho_0^2}}
g^N_0\Big(\frac{\rho\rho_0}{\rho^2+\rho_0^2}\Big)\simeq 0.16.
\label{B}
\eeq

\section{Boundary conditions at the bottom of the viscous layer}
\label{Boundary conditions}
The derivation of Eq. (\ref{eq3}) is straightforward and is omitted. We concentrate
on continuity of normal stress.
We need first to enforce continuity
of the normal velocity:
\beq
\PhiU_+\ex^{-kh}-\PhiU_-\ex^{kh}-
\im(\AU_+\ex^{-\alpha_kh}+\AU_-\ex^{\alpha_kh})=\PhiU_w\sinh[k(H-h)].
\label{cacca}
\eeq
Continuity of normal stress gives
\beq
2\nu\partial_z U_z|_{z=-h^+}-P/\varrho=-P_w/\varrho.
\label{cacca1}
\eeq
The first of Eq. (\ref{potential pressure}) allows us to write
\beq
P=\varrho\{-\im\omega(\PhiU_+\ex^{-kh}+\PhiU_-\ex^{kh})
+\frac{\im gk}{\omega}\PhiU_w\sinh[k(H-h)]\}
\label{cacca2}
\eeq
and
\beq
P_w=\varrho_w\PhiU_w\{-\im\omega\cosh(k(H-h))+\frac{\im gk}{\omega}\sinh[k(H-h)]\}.
\label{cacca3}
\eeq
Substituting Eqs. (\ref{cacca2}) and (\ref{cacca3}) into Eq. (\ref{cacca1}), 
and passing to dimensionless variables, we get
\begin{eqnarray}
&&\hatrho(\im-2\hat\nu\hat k^2)(\PhiU_+\ex^{-\hat k\hat h}+\PhiU_-\ex^{\hat k\hat h})
+2\im\hatrho\hat\nu^{1/2}\hat\alpha\hat k( \AU_+\ex^{-\hat\alpha_k\psi}-
 \AU_-\ex^{\hat\alpha_k\psi})
\nonumber
\\
&&-\im\{\qHh-(1-\hatrho)\hat k\}\PhiU_w\sinh[\hat k(\hat H-\hat h)]=0,
\end{eqnarray}
where $\qHh=1/\tanh[k(H-h)]$.
We can eliminate $\PhiU_w$ using Eq. (\ref{cacca}), to obtain
\begin{eqnarray}
&&\{\im[\hatrho-\qHh+(1-\hatrho)\hat k]-2\hatrho\hat\nu\hat k^2\}\PhiU_+\ex^{-\hat k\hat h}
+\{\im[\hatrho+\qHh-(1-\hatrho)\hat k]-2\hatrho\hat\nu\hat k^2\}\PhiU_-\ex^{\hat k\hat h}
\nonumber
\\
&&+[(1-\hatrho)\hat k-\qHh+2\im\hatrho\hat\nu^{1/2}\hat\alpha\hat k] \AU_+
\ex^{-\hat\alpha_k\psi}
\nonumber
\\
&&+[(1-\hatrho)\hat k-\qHh-2\im\hatrho\hat\nu^{1/2}\hat\alpha\hat k] \AU_-
\ex^{\hat\alpha_k\psi}
=0,
\end{eqnarray}
that is Eq. (\ref{eq4}).

\bibliography{panbib}

\begin{thebibliography}{29}
\providecommand{\natexlab}[1]{#1}
\providecommand{\url}[1]{\texttt{#1}}
\expandafter\ifx\csname urlstyle\endcsname\relax
  \providecommand{\doi}[1]{doi: #1}\else
  \providecommand{\doi}{doi: \begingroup \urlstyle{rm}\Url}\fi

\bibitem[Bennetts and Squire(2009)]{bennetts09}
LG~Bennetts and VA~Squire.
\newblock Wave scattering by multiple rows of circular ice floes.
\newblock \emph{Journal of Fluid Mechanics}, 639:\penalty0 213--238, 2009.

\bibitem[Brekke and Solberg(2005)]{brekke05}
Camilla Brekke and Anne~HS Solberg.
\newblock Oil spill detection by satellite remote sensing.
\newblock \emph{Remote sensing of environment}, 95:\penalty0 1--13, 2005.

\bibitem[De~Carolis and Desiderio(2002)]{decarolis02}
Giacomo De~Carolis and Daniela Desiderio.
\newblock Dispersion and attenuation of gravity waves in ice: a two-layer
  viscous fluid model with experimental data validation.
\newblock \emph{Physics Letters A}, 305:\penalty0 399--412, 2002.

\bibitem[de~Carolis et~al.(2005)de~Carolis, Olla, and Pignagnoli]{decarolis05}
Giacomo de~Carolis, Piero Olla, and Luca Pignagnoli.
\newblock Effective viscosity of grease ice in linearized gravity waves.
\newblock \emph{Journal of Fluid Mechanics}, 535:\penalty0 369--381, 2005.

\bibitem[Doble et~al.(2015)Doble, De~Carolis, Meylan, Bidlot, and
  Wadhams]{doble15}
Martin~J Doble, Giacomo De~Carolis, Michael~H Meylan, Jean-Raymond Bidlot, and
  Peter Wadhams.
\newblock Relating wave attenuation to pancake ice thickness, using field
  measurements and model results.
\newblock \emph{Geophysical Research Letters}, 42:\penalty0 4473--4481, 2015.

\bibitem[Fingas and Hollebone(2003)]{fingas03}
MF~Fingas and BP~Hollebone.
\newblock Review of behaviour of oil in freezing environments.
\newblock \emph{Marine Pollution Bulletin}, 47:\penalty0 333--340, 2003.

\bibitem[Foldy(1945)]{foldy45}
Leslie~L Foldy.
\newblock The multiple scattering of waves. i. general theory of isotropic
  scattering by randomly distributed scatterers.
\newblock \emph{Physical Review}, 67:\penalty0 107, 1945.

\bibitem[Jackson(1999)]{jackson}
John~David Jackson.
\newblock \emph{Classical electrodynamics}.
\newblock Wiley, 1999.

\bibitem[Keller(1998)]{keller98}
Joseph~B Keller.
\newblock Gravity waves on ice-covered water.
\newblock \emph{Journal of Geophysical Research: Oceans}, 103:\penalty0
  7663--7669, 1998.

\bibitem[Kohout and Meylan(2008)]{kohout08}
AL~Kohout and MH~Meylan.
\newblock An elastic plate model for wave attenuation and ice floe breaking in
  the marginal ice zone.
\newblock \emph{Journal of Geophysical Research: Oceans}, 113:\penalty0 C09016,
  2008.

\bibitem[Kohout et~al.(2014)Kohout, Williams, Dean, and Meylan]{kohout14}
AL~Kohout, MJM Williams, SM~Dean, and MH~Meylan.
\newblock Storm-induced sea-ice breakup and the implications for ice extent.
\newblock \emph{Nature}, 509:\penalty0 604--607, 2014.

\bibitem[Lamb(1932)]{lamb}
Horace Lamb.
\newblock \emph{Hydrodynamics}.
\newblock Cambridge university press, 1932.

\bibitem[Longuet-Higgins(1953)]{longuet53}
Michael~S Longuet-Higgins.
\newblock Mass transport in water waves.
\newblock \emph{Philosophical Transactions of the Royal Society of London A:
  Mathematical, Physical and Engineering Sciences}, 245:\penalty0 535--581,
  1953.

\bibitem[Meylan(2002)]{meylan02}
Michael~H Meylan.
\newblock Wave response of an ice floe of arbitrary geometry.
\newblock \emph{Journal of Geophysical Research: Oceans}, 107:\penalty0 3005,
  2002.

\bibitem[Mosig et~al.(2015)Mosig, Montiel, and Squire]{mosig15}
Johannes~EM Mosig, Fabien Montiel, and Vernon~A Squire.
\newblock Comparison of viscoelastic-type models for ocean wave attenuation in
  ice-covered seas.
\newblock \emph{Journal of Geophysical Research: Oceans}, 120:\penalty0
  6072--6090, 2015.

\bibitem[Newyear and Martin(1999)]{newyear99}
Karl Newyear and Seelye Martin.
\newblock Comparison of laboratory data with a viscous two-layer model of wave
  propagation in grease ice.
\newblock \emph{Journal of Geophysical Research: Oceans}, 104:\penalty0
  7837--7840, 1999.

\bibitem[Shen and Squire(1998)]{shen98}
Hayley~H Shen and Vernon~A Squire.
\newblock \emph{Wave damping in compact pancake ice fields due to interactions
  between pancakes}, volume~74 of \emph{Antarctic Science Series}, pages
  343--351.
\newblock American Geophysical Union, 1998.

\bibitem[Squire(2007)]{squire07}
VA~Squire.
\newblock Of ocean waves and sea-ice revisited.
\newblock \emph{Cold Regions Science and Technology}, 49:\penalty0 110--133,
  2007.

\bibitem[Squire and Williams(2008)]{squire08}
VA~Squire and TD~Williams.
\newblock Wave propagation across sea-ice thickness changes.
\newblock \emph{Ocean Modelling}, 21:\penalty0 1--11, 2008.

\bibitem[Squire et~al.(1995)Squire, Dugan, Wadhams, Rottier, and Liu]{squire95}
Vernon~A Squire, John~P Dugan, Peter Wadhams, Philip~J Rottier, and Antony~K
  Liu.
\newblock Of ocean waves and sea ice.
\newblock \emph{Annual Review of Fluid Mechanics}, 27:\penalty0 115--168, 1995.

\bibitem[Wadhams et~al.(2002)Wadhams, Parmiggiani, and De~Carolis]{wadhams02}
P~Wadhams, F~Parmiggiani, and G~De~Carolis.
\newblock The use of sar to measure ocean wave dispersion in frazil-pancake
  icefields.
\newblock \emph{Journal of physical oceanography}, 32:\penalty0 1721--1746,
  2002.

\bibitem[Wadhams et~al.(2004)Wadhams, Parmiggiani, De~Carolis, Desiderio, and
  Doble]{wadhams04}
P~Wadhams, FF~Parmiggiani, G~De~Carolis, D~Desiderio, and MJ~Doble.
\newblock Sar imaging of wave dispersion in antarctic pancake ice and its use
  in measuring ice thickness.
\newblock \emph{Geophysical research letters}, 31:\penalty0 L15305, 2004.

\bibitem[Wadhams(1973)]{wadhams73}
Peter Wadhams.
\newblock Attenuation of swell by sea ice.
\newblock \emph{Journal of Geophysical Research}, 78:\penalty0 3552--3563,
  1973.

\bibitem[Wadhams and Holt(1991)]{wadhams91}
Peter Wadhams and Benjamin Holt.
\newblock Waves in frazil and pancake ice and their detection in seasat
  synthetic aperture radar imagery.
\newblock \emph{Journal of Geophysical Research: Oceans}, 96:\penalty0
  8835--8852, 1991.

\bibitem[Wadhams et~al.(1988)Wadhams, Squire, Goodman, Cowan, and
  Moore]{wadhams88}
Peter Wadhams, Vernon~A Squire, Dougal~J Goodman, Andrew~M Cowan, and Stuart~C
  Moore.
\newblock The attenuation rates of ocean waves in the marginal ice zone.
\newblock \emph{Journal of Geophysical Research: Oceans}, 93:\penalty0
  6799--6818, 1988.

\bibitem[Wang and Shen(2010{\natexlab{a}})]{wang10}
Ruixue Wang and Hayley~H Shen.
\newblock Gravity waves propagating into an ice-covered ocean: A viscoelastic
  model.
\newblock \emph{Journal of Geophysical Research: Oceans}, 115:\penalty0 C06024,
  2010{\natexlab{a}}.

\bibitem[Wang and Shen(2010{\natexlab{b}})]{wang10a}
Ruixue Wang and Hayley~H Shen.
\newblock Experimental study on surface wave propagating through a
  grease--pancake ice mixture.
\newblock \emph{Cold Regions Science and Technology}, 61:\penalty0 90--96,
  2010{\natexlab{b}}.

\bibitem[Wang and Shen(2011)]{wang11}
Ruixue Wang and Hayley~H Shen.
\newblock A continuum model for the linear wave propagation in ice-covered
  oceans: An approximate solution.
\newblock \emph{Ocean Modelling}, 38:\penalty0 244--250, 2011.

\bibitem[Weber(1987)]{weber87}
Jan~Erik Weber.
\newblock Wave attenuation and wave drift in the marginal ice zone.
\newblock \emph{Journal of physical oceanography}, 17:\penalty0 2351--2361,
  1987.

\end{thebibliography}

\end{document}